\documentclass[aps,prb,reprint,groupedaddress]{revtex4-2}

\usepackage{graphicx}
\usepackage{amsmath,amssymb}
\usepackage{mathrsfs}

\begin{document}

%\preprint{}


\title{Higgs-mode-induced instability and kinetic inductance in strongly dc-biased dirty-limit superconductors}


\author{Takayuki Kubo}
\email[]{kubotaka@post.kek.jp}
%\thanks{}
\affiliation{High Energy Accelerator Research Organization (KEK), Tsukuba, Ibaraki 305-0801, Japan}
\affiliation{The Graduate University for Advanced Studies (Sokendai), Hayama, Kanagawa 240-0193, Japan}


%\date{\today}

\begin{abstract}
A perturbative ac field superposed on a dc bias (\( J_b \)) is known to excite the Higgs mode in superconductors.  
The dirty limit, where disorder enhances the Higgs resonance, provides an ideal setting for this study and is also relevant to many superconducting devices operating under strong dc biases.  
However, the effects of a strong dc bias near the depairing current in dirty-limit superconductors remain largely unexplored, despite their significance in both fundamental physics and applications.  
In this paper, we derive a general formula for the complex conductivity of disordered superconductors under an \textit{arbitrary} dc bias using the Keldysh-Usadel theory of nonequilibrium superconductivity.  
This formula is relatively simple, making it more accessible to a broader research community.  
Our analysis reveals that in a strongly dc-biased dirty-limit superconductor, the Higgs mode induces an instability in the homogeneous superflow within a specific frequency window, making the high-current-carrying state vulnerable to ac perturbations.  
This instability, which occurs exclusively in the \( {\rm ac} \parallel {\rm dc} \) configuration, leads to a non-monotonic dependence of kinetic inductance on frequency and bias strength.  
By carefully tuning the dc bias and the frequency of the ac perturbation, the kinetic inductance can be enhanced by nearly two orders of magnitude.  
In the weak dc bias regime, our formula recovers the well-known quadratic dependence, \( L_k \propto 1+ C(J_b/J_{\rm dp})^2 \), with coefficients \( C=0.409 \) for \( {\rm ac} \parallel {\rm dc} \) and \( C=0.136 \) for \( {\rm ac} \perp {\rm dc} \), where \( J_{\rm dp} \) is the equilibrium depairing current density.  
These findings establish a robust theoretical framework for dc-biased superconducting systems and suggest that Higgs mode physics could be exploited in the design and optimization of superconducting detectors.  
Moreover, they may lead to a yet-to-be-explored detector concept based on Higgs mode physics.  
\end{abstract}

\maketitle


%%%%%%%%%%%%%
\section{Introduction}\label{intro}
%%%%%%%%%%%%%

A conventional \( s \)-wave superconductor under a dc bias is a fundamental yet intricate physical system that underpins various applications, including kinetic inductance detectors~\cite{Zmuidzinas}, traveling-wave parametric amplifiers~\cite{Visser}, neutron detector~\cite{Shishido}, single-photon detectors~\cite{Zadeh}, superconducting diodes~\cite{Ando}, and the fundamental study of superconducting rf cavities for particle accelerators~\cite{Gurevich_Review}.  
Although the behavior of a dc-biased superconductor may initially seem straightforward, the introduction of an ac field parallel to the bias (\({\rm ac} \parallel {\rm dc}\)) gives rise to highly nontrivial physical phenomena, where the Higgs mode plays a pivotal role~\cite{Moor, Nakamura, Jujo, 2024_Kubo}.

\begin{figure}[tb]
   \begin{center}
   \includegraphics[width=0.95\linewidth]{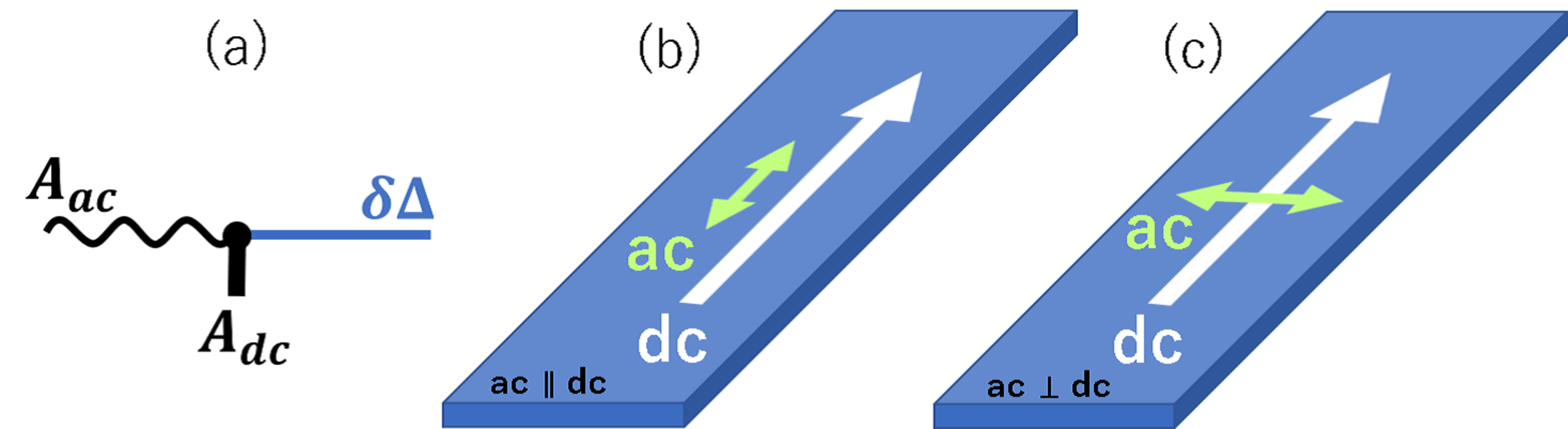}
   \end{center}\vspace{0 cm}
   \caption{
(a) Schematic illustration of the Higgs mode interaction with the electromagnetic field.  
When an ac field is superposed on a bias dc in parallel, the bias dc acts as a tuning knob, enabling the Higgs mode to respond linearly and be amplified.  
(b, c) Narrow superconducting thin film under a bias dc with a superposed ac field in the  
(b) \( {\rm ac} \parallel {\rm dc} \) and (c) \( {\rm ac} \perp {\rm dc} \) configurations.  
In the \( {\rm ac} \parallel {\rm dc} \) case, the Higgs mode responds linearly to the ac perturbation, whereas in the \( {\rm ac} \perp {\rm dc} \) case, no such response occurs.  
   }\label{fig1}
\end{figure}

The Higgs mode, \(\delta \Delta(t)\), corresponds to oscillations in the amplitude of the superconducting order parameter \(\Delta\) (see, e.g., Refs.~\cite{Anderson, Volkov_Kogan, Shimano_review, Tsuji_review, 2013_Matsunaga, 2014_Matsunaga, Tsuji_Aoki, 2018_Jujo, Silaev}). 
It is well known that the Higgs mode couples to electromagnetic fields through a term of the form \( {\bf A} \cdot {\bf A}  \delta \Delta \) [Fig.~\ref{fig1}(a)], similar to how the Higgs boson \( h \) in particle physics couples to the \( Z \) boson via a \( Z Z h \) interaction.  
As a result, the Higgs mode typically appears in nonlinear responses and requires strong electromagnetic irradiation to be observed~\cite{2013_Matsunaga, 2014_Matsunaga}.  
However, when an ac field is superposed on a dc bias, the coupling term becomes \( {\bf A}_{\rm dc} \cdot {\bf A} \delta \Delta \)~\cite{Moor, Nakamura, Jujo, 2024_Kubo}, allowing the Higgs mode to respond \textit{linearly} to the electromagnetic field in the \({\rm ac} \parallel {\rm dc}\) configuration [Fig.~\ref{fig1}(b)].  
This effect was first identified by Moor et al.~\cite{Moor}, who predicted that the Higgs resonance would appear at \( \hbar \omega = 2 \Delta \) for \( {\rm ac} \parallel {\rm dc} \).  
This prediction was later confirmed experimentally by Nakamura et al.~\cite{Nakamura}.  
For the \( {\rm ac} \perp {\rm dc} \) configuration [Fig.~\ref{fig1}(c)], however, this coupling vanishes.  
It is worth noting that Budzinski et al.~\cite{Budzinski} seemingly observed the Higgs mode under a dc bias nearly half a century ago using aluminum samples with varying mean free paths (see also the discussion in Ref.~\cite{2024_Kubo}).

The pioneering work by Moor et al.~\cite{Moor} mainly considered weak dc biases, treating both the ac field and dc bias as small perturbations.  
However, the \textit{nonperturbative} effects of a strong dc bias on the Higgs resonance and its broader implications remained largely unexplored.  
To address this, Jujo~\cite{Jujo} developed a general theoretical framework based on the Keldysh-Eilenberger formalism to compute the surface resistance of superconductors with arbitrary mean free path and bias dc strength.  
Expanding upon this framework, Ref.~\cite{2024_Kubo} computed the complex conductivity over a wide range of bias current strengths, ac-perturbation frequencies, and temperatures, revealing that the Higgs mode contribution is not only significant near \( \hbar \omega \simeq 2\Delta \) but also at \( \hbar \omega \ll \Delta \), a frequency range directly relevant to superconducting devices.

Among the key findings of Ref.~\cite{2024_Kubo}, one of the most striking results is the profound impact of the Higgs mode on the bias-dependent kinetic inductance \( L_k(J_b) \).  
The coefficient \( C \) in the low-bias expansion:  
\begin{eqnarray}
L_k(J_b) = L_k(0) \biggl[ 1 + C \biggl( \frac{J_b}{J_{\rm dp}} \biggr)^2 + \dots \biggr], \label{Lk_expansion}
\end{eqnarray}
where \( J_{\rm dp} \) is the equilibrium depairing current density, was numerically determined for various mean free paths and temperatures.  
For the \( {\rm ac} \parallel {\rm dc} \) configuration, \( C \) was found to approach \( C \simeq 0.4 \) as the mean free path decreases.  
On the other hand, \( C \) for the \( {\rm ac} \perp {\rm dc} \) configuration, obtained by omitting nonequilibrium corrections due to the ac perturbation, approaches \( C \simeq 0.14 \).  

Notably, these numerical results align with previous analytical studies of \( L_k(J_b) \), where \( C \) was derived using the equilibrium Usadel equation combined with one of the following phenomenological assumptions about nonequilibrium effects:  
(i) The \textit{oscillating \( n_s \) assumption}, in which the superfluid density (\( n_s \)) oscillates in sync with the alternating current, leading to \( C(T \to 0) \simeq 0.409 \)~\cite{2020_Kubo, 2020_Kubo_erratum}.  
(ii) The \textit{frozen \( n_s \) assumption}, in which the superfluid density remains fixed at its equilibrium value set by the dc bias, yielding \( C(T \to 0) \simeq 0.136 \)~\cite{2020_Kubo}.  

These phenomenological assumptions~\cite{2020_Kubo, 2020_Kubo_erratum, Anlage, Clem_Kogan} were historically motivated by the idea that if the ac frequency is lower than the inverse relaxation time of the superfluid density, \( n_s \) will oscillate along with the ac; conversely, if the ac frequency is higher, \( n_s \) will remain fixed at its equilibrium value determined by the strength of the bias dc.  
Thus, these assumptions were referred to as the \textit{slow experiment} and \textit{fast experiment} assumptions.  
However, it is now evident that the distinction between these regimes is not determined by whether the ac frequency is slow or fast, but rather by whether the Higgs mode is excited (\({\rm ac} \parallel {\rm dc}\)) or remains inactive (\({\rm ac} \perp {\rm dc}\)).

Despite these advancements, initiated by the pioneering work of Moor et al.~\cite{Moor}, we can expect that some intriguing phenomena remain hidden. 
These would arise in a dirty-limit superconductor under a strong bias dc comparable to \( J_{\rm dp} \), because the Higgs resonance becomes increasingly pronounced as the mean free path shortens and the dc bias is strengthened~\cite{Jujo, 2024_Kubo}.  
This provides an ideal setting where Higgs mode physics is most prominent.  
Coincidentally, this is also the operating regime of various superconducting applications.

There are several challenges and open questions:  
(i) The complex conductivity formula for a dirty-limit superconductor under an arbitrary bias dc, which is crucial for investigating this regime, has yet to be established.
While a more general expression covering the entire range from the dirty to clean limit has been derived using the Keldysh-Eilenberger formalism~\cite{Jujo, 2024_Kubo}, a specialized formula for the dirty limit is expected to be significantly simpler.
Such a formulation would not only provide deeper physical insight but also make the theoretical framework more accessible to a broader range of researchers.
(ii) Does the stability of the homogeneous superflow persist in the presence of a strong Higgs resonance (i.e., oscillations of the superfluid density), particularly as the mean free path decreases and the dc bias increases?  
(iii) In the previous study~\cite{2024_Kubo}, the coefficient \( C \) was numerically computed, but its exact analytical value in the dirty limit remains undetermined.  
Do these values exactly coincide with those previously obtained under the oscillating and frozen superfluid density assumptions?  
(iv) How does bias-dependent kinetic inductance behave under strong dc bias in a dirty-limit superconductor, where the Higgs resonance is pronounced, and can these effects be leveraged for designing and optimizing superconducting devices?

In this paper, we address these outstanding issues using the Keldysh-Usadel theory of nonequilibrium superconductivity.  
The paper is organized as follows:  
In Section~II, we derive the complex conductivity formula for a disordered superconductor subjected to an ac perturbation superposed on a bias dc of arbitrary strength.  
In Section~III, we apply this formula to investigate the Higgs mode contribution to the complex conductivity and examine the stability of the homogeneous superflow.  
In Section~IV, we analyze the bias-dependent kinetic inductance.  
Finally, Section~V summarizes our findings and discusses their broader implications, including superconducting device applications.  
Readers unfamiliar with the theoretical methods used in this paper may skip Sections~II, III, and IV and proceed directly to this section.

%%%%%%%%%%%%%%%%%%%
%%%%%%%%%%%%%%%%%%%
\section{Formulation} \label{formulation}
%%%%%%%%%%%%%%%%%%%
%%%%%%%%%%%%%%%%%%%

The purpose of this section is to derive the complex conductivity formula for a disordered superconductor under an arbitrary bias dc strength, starting from the Keldysh-Usadel equations.

%%%%%%%%%%%%%%%%%%%
%%%%%%%%%%%%%%%%%%%
\subsection{Keldysh-Usadel equations}
%%%%%%%%%%%%%%%%%%%
%%%%%%%%%%%%%%%%%%%

We examine the behavior of a narrow thin-film superconductor, subjected to a bias dc current of arbitrary magnitude that flows parallel to the strip. Superimposed on this is a weak electromagnetic field. The formalism used is the Keldysh-Usadel quasiclassical Green's functions, which is a well-established approach for analyzing nonequilibrium superconductivity in the dirty limit (see, e.g., Refs.~\cite{Kopnin, Rammer, Sauls}).
In this framework, the quasiclassical Green's functions are expressed in a mixed Fourier representation with retarded, advanced, and Keldysh components, denoted by $\hat{g}^{R, A, K}(\epsilon, t)$. These are $2 \times 2$ matrix functions in Nambu space. The governing equations for the retarded (R) and advanced (A) components of the Green's functions are given by
\begin{eqnarray}
&&i\hbar D \hat{\partial} \circ \big\{ \hat{g}^{r} \circ (\hat{\partial} \circ \hat{g}^{r}) \big\}
=[\epsilon \hat{\tau}_3 + \hat{\Delta} , \hat{g}^{r}]_{\circ} , \label{RA1} \\
&&\hat{g}^R \circ \hat{g}^R = \hat{g}^A \circ \hat{g}^A = 1 , \label{normRA1}
\end{eqnarray}
where $\hat{\tau}_3$ is the third Pauli matrix and $\hat{\Delta}$ represents the superconducting gap. The Keldysh component $\hat{g}^K$ satisfies a similar equations:
\begin{eqnarray}
&&i\hbar D \hat{\partial} \circ \big\{ \hat{g}^{R} \circ (\hat{\partial} \circ \hat{g}^{K}) + \hat{g}^K \circ  (\hat{\partial} \circ \hat{g}^{A})  \big\} %\nonumber \\
=  [\epsilon \hat{\tau}_3 + \hat{\Delta} , \hat{g}^{K}]_{\circ} , \nonumber \\ \label{K1} \\
&&\hat{g}^R \circ \hat{g}^K + \hat{g}^K \circ  \hat{g}^A=0 . \label{normK1}
\end{eqnarray}
The $\circ$ product here refers to the Moyal product, defined as
$X \circ Y = \exp\left[(i\hbar/2)\left(\partial_\epsilon^X \partial_t^Y - \partial_\epsilon^Y \partial_t^X\right)\right] XY$,
while $[X, Y]_{\circ}$ represents the commutator in this context. Under the influence of a gauge-invariant superfluid momentum $\hbar {\bf q} = \hbar \nabla \chi -2e{\bf A}$, the operator $\hat{\partial} \circ$ reduces to:
$\hat{\partial} \circ \hat{g}^{R, A, K} =  (i/2) [ \hat{\tau}_3 {\bf q} , \hat{g}]_{\circ}$. 
The gap equation governing the superconducting order parameter $\Delta$ is derived from the Keldysh component and is expressed as
\begin{eqnarray}
\hat{\Delta}(t) = -\frac{{\mathscr G}}{8}\int d\epsilon {\rm Tr} [(-i\tau_2) \hat{g}^K(\epsilon, t)]  , \label{gap1}
\end{eqnarray}
where ${\mathscr G}$ is the BCS coupling constant. 
The current density ${\bf J}(t)$ in the superconductor is given by
\begin{eqnarray}
&& {\bf J}(t) = -i \frac{\sigma_n}{e}\int \!\!d\epsilon {\bf S}(\epsilon, t) , \\
&& {\bf S}(\epsilon, t)= \frac{i}{8} {\rm Tr} \bigl[ \hat{\tau}_3 
\bigl\{ \hat{g}^{R} \circ (\hat{\partial} \circ \hat{g}^{K}) + \hat{g}^K \circ  (\hat{\partial} \circ \hat{g}^{A}) \bigr\} \bigr]  ,  \label{S1} 
\end{eqnarray}
where $\sigma_n=2e^2 N_0 D$ is the normal conductivity, $N_0$ is the normal density of states at the Fermi level and $D$ is the diffusion constant.

Eqs.~(\ref{RA1})-(\ref{S1}) form the general framework of the Keldysh-Usadel formalism for nonequilibrium superconductivity. 
From this framework, we analyze the linear response to a weak electromagnetic field in the presence of a dc bias of arbitrary strength.

%%%%%%%%%%%%%%%%%%%
\subsection{Linear response to a weak electromagnetic field in the presence of a bias dc of arbitrary strength}
%%%%%%%%%%%%%%%%%%%

We define the total momentum as $\mathbf{q} = \mathbf{q}_b + \delta \mathbf{q}(t)$, where $\mathbf{q}_b$ represents the constant bias momentum due to the dc component. The strength of $\mathbf{q}_b$ can vary from zero up to the depairing momentum. The term $\delta \mathbf{q}(t)$ denotes the time-dependent perturbation applied to the system.

In this framework, the Green's functions $\hat{g}^{R,A,K}(\epsilon, t)$ are decomposed as $\hat{g}^{R,A,K}(\epsilon, t) = \hat{g}^{R,A,K}_b(\epsilon) + \delta \hat{g}^{R,A,K}(\epsilon, t)$, where $\hat{g}^{R,A,K}_b(\epsilon)$ represents the equilibrium component in the presence of the dc bias, and $\delta \hat{g}^{R,A,K}(\epsilon, t)$ accounts for the perturbative response. Similarly, the superconducting order parameter is expressed as $\hat{\Delta}(t) = \hat{\Delta}_b + \delta \hat{\Delta}(t)$, with $\hat{\Delta}_b$ representing the bias-induced equilibrium value and $\delta \hat{\Delta}(t)$ represents the time-dependent fluctuation of the order parameter, the Higgs mode. This decomposition allows us to reformulate Eqs.~(\ref{RA1})-(\ref{S1}) in terms of orders of approximation.

The zeroth-order equations, which describe the equilibrium Green's functions for an arbitrary strength of dc bias, take the form: 
\begin{eqnarray}
&&-i (s/2) \bigl\{ \hat{\tau}_{3} \hat{g}^{r}_{b}(\epsilon) \hat{\tau}_{3} \hat{g}^{r}_{b}(\epsilon) - \hat{g}^{r}_{b}(\epsilon) \hat{\tau}_{3} \hat{g}^{r}_{b}(\epsilon) \hat{\tau}_{3} \bigr\} \nonumber \\
&&= \left[ \epsilon \hat{\tau}_{3} + \hat{\Delta}_{b} , \hat{g}^{r}_{b}(\epsilon)  \right]  
\hspace{1cm} (r=R, A) , \label{0thRA} \\
&& \hat{g}_b^K(\epsilon) = \bigl\{ \hat{g}_b^R(\epsilon) -\hat{g}_b^A(\epsilon) \bigr\} \mathcal{T}(\epsilon) , \\
&&\Delta_b   
=\frac{{\mathscr G}}{2} \int \!\! d\epsilon F_b(\epsilon) \mathcal{T}(\epsilon)    , \label{gap_bias}
\end{eqnarray}
with the normalization condition $\hat{g}^{r}_{b}(\epsilon) \hat{g}^{r}_{b}(\epsilon) = 1$ and 
$\hat{g}^{R}_{b}(\epsilon) \hat{g}^{K}_{b}(\epsilon)  + \hat{g}^{K}_{b}(\epsilon) \hat{g}^{A}_{b}(\epsilon) =0$, 
where 
\begin{eqnarray}
s := \frac{\hbar D}{2} q_b^2 ,
\end{eqnarray}
represents the strength of the bias superflow, and  $\mathcal{T}(\epsilon) := \tanh \left( \epsilon / 2kT \right) = 1 - 2f_{\rm FD}(\epsilon)$ is the equilibrium distribution function, with $f_{\rm FD}(\epsilon)$ representing the Fermi-Dirac distribution. 
These equations describe the equilibrium Green's functions in the presence of a bias dc and are applicable to bias dc of any strength.

At the first-order approximation with respect to the weak ac field, the equations governing the $R$ and $A$ components of the perturbed Green's functions, $\delta \hat{g}^r(\epsilon, \omega)$ ($r = R, A$), in Fourier space are given by: 
\begin{eqnarray}
&&-i (s/2) \bigl\{ 
\hat{\tau}_{3} \hat{g}^{r}_{b}(\epsilon_+) \hat{\tau}_{3} \delta \hat{g}^{r}(\epsilon, \omega) 
- \hat{g}^{r}_{b}(\epsilon_+) \hat{\tau}_{3} \delta \hat{g}^{r}(\epsilon, \omega) \hat{\tau}_{3} \nonumber \\
&&+ \hat{\tau}_{3} \delta \hat{g}^{r}(\epsilon, \omega) \hat{\tau}_{3} \hat{g}^{r}_{b}(\epsilon_-) 
- \delta \hat{g}^{r}(\epsilon, \omega) \hat{\tau}_{3} \hat{g}^{r}_{b}(\epsilon_-) \hat{\tau}_{3} \bigr\} \nonumber \\
&&- i(\delta W/2)  \big\{ 
\hat{\tau}_{3} \hat{g}^{r}_{b}(\epsilon_+) \hat{\tau}_{3} \hat{g}^{r}_{b}(\epsilon_-) \nonumber \\
&&- \hat{g}^{r}_{b}(\epsilon_+) \hat{\tau}_{3} \hat{g}^{r}_{b}(\epsilon_-) \hat{\tau}_{3} 
+ \hat{\tau}_{3} \hat{g}^{r}_{b}(\epsilon_-) \hat{\tau}_{3} \hat{g}^{r}_{b}(\epsilon_-) \nonumber \\
&&- \hat{g}^{r}_{b}(\epsilon_+) \hat{\tau}_{3} \hat{g}^{r}_{b}(\epsilon_+) \hat{\tau}_{3} \bigr\} 
=   \epsilon_+ \hat{\tau}_{3} \delta \hat{g}^{r}(\epsilon, \omega) - \delta \hat{g}^{r}(\epsilon, \omega) \hat{\tau}_{3} \epsilon_-  \nonumber  \\
&&+ [ \hat{\Delta}_{b} , \delta \hat{g}^{r}(\epsilon, \omega) ] 
+ \delta\hat{\Delta}(\omega) \hat{g}^{r}_{b}(\epsilon_-) - \hat{g}^{r}_{b}(\epsilon_+) \delta\hat{\Delta}(\omega) , \label{1st_order_RA}
\end{eqnarray}
with the normalization condition
\begin{eqnarray}
\hat{g}^{r}_{b}(\epsilon_+) \delta \hat{g}^{r}(\epsilon, \omega) + \delta \hat{g}^{r}(\epsilon, \omega) \hat{g}^{r}_{b}(\epsilon_-) =0 , \label{1st_order_norm_RA}
\end{eqnarray}
where $\delta W=(\hbar D/2) {\bf q}_b \cdot \delta {\bf q}_{\omega}$, 
$\delta {\bf q}_{\omega}=\int \delta {\bf q} (t) e^{i\omega t} dt$ is Fourier transform of the ac field $\delta {\bf q}(t)$, and $\epsilon_\pm = \epsilon \pm \hbar \omega/2$. 
The equation for the Keldysh component in Fourier space, $\delta \hat{g}^K(\epsilon, \omega)$, is given by
\begin{eqnarray}
&&-i (s/2)  \bigl[ 
\hat{\tau}_{3} \hat{g}^{R}_{b}(\epsilon_+) \hat{\tau}_{3} \delta \hat{g}^{K}(\epsilon, \omega) 
- \hat{g}^{R}_{b}(\epsilon_+) \hat{\tau}_{3} \delta \hat{g}^{K}(\epsilon, \omega) \hat{\tau}_{3} \nonumber \\
&&+ \hat{\tau}_{3} \delta \hat{g}^{R}(\epsilon, \omega) \hat{\tau}_{3} \hat{g}^{K}_{b}(\epsilon_-) 
- \delta \hat{g}^{R}(\epsilon, \omega) \hat{\tau}_{3} \hat{g}^{K}_{b}(\epsilon_-) \hat{\tau}_{3} \nonumber \\
&& + \hat{\tau}_{3} \hat{g}^{K}_{b}(\epsilon_+) \hat{\tau}_{3} \delta \hat{g}^{A}(\epsilon, \omega) 
- \hat{g}^{K}_{b}(\epsilon_+) \hat{\tau}_{3} \delta \hat{g}^{A}(\epsilon, \omega) \hat{\tau}_{3} \nonumber \\
&&+ \hat{\tau}_{3} \delta \hat{g}^{K}(\epsilon, \omega) \hat{\tau}_{3} \hat{g}^{A}_{b}(\epsilon_-) 
- \delta \hat{g}^{K}(\epsilon, \omega) \hat{\tau}_{3} \hat{g}^{A}_{b}(\epsilon_-) \hat{\tau}_{3} 
\bigr] \nonumber \\
&&- i(\delta W/2) \big[ \hat{\tau}_{3} \hat{g}^{R}_{b}(\epsilon_+) \hat{\tau}_{3} \hat{g}^{K}_{b}(\epsilon_-) 
- \hat{g}^{R}_{b}(\epsilon_+) \hat{\tau}_{3} \hat{g}^{K}_{b}(\epsilon_-) \hat{\tau}_{3} \nonumber \\
&&+ \hat{\tau}_{3} \hat{g}^{R}_{b}(\epsilon_-) \hat{\tau}_{3} \hat{g}^{K}_{b}(\epsilon_-) 
- \hat{g}^{R}_{b}(\epsilon_+) \hat{\tau}_{3} \hat{g}^{K}_{b}(\epsilon_+) \hat{\tau}_{3} \nonumber \\
&&+ \hat{\tau}_{3} \hat{g}^{K}_{b}(\epsilon_+) \hat{\tau}_{3} \hat{g}^{A}_{b}(\epsilon_-) 
- \hat{g}^{K}_{b}(\epsilon_+) \hat{\tau}_{3} \hat{g}^{A}_{b}(\epsilon_-) \hat{\tau}_{3} \nonumber \\
&&+ \hat{\tau}_{3} \hat{g}^{K}_{b}(\epsilon_-) \hat{\tau}_{3} \hat{g}^{A}_{b}(\epsilon_-) 
- \hat{g}^{K}_{b}(\epsilon_+) \hat{\tau}_{3} \hat{g}^{A}_{b}(\epsilon_+) \hat{\tau}_{3} \bigr] \nonumber \\
&& =   \epsilon_+ \hat{\tau}_{3} \delta \hat{g}^{K}(\epsilon, \omega) - \delta \hat{g}^{K}(\epsilon, \omega) \hat{\tau}_{3} \epsilon_-  \nonumber  \\
&&+ [ \hat{\Delta}_{b} , \delta \hat{g}^{K}(\epsilon, \omega) ] 
+ \delta\hat{\Delta}(\omega) \hat{g}^{K}_{b}(\epsilon_-) - \hat{g}^{K}_{b}(\epsilon_+) \delta\hat{\Delta}(\omega) . \label{1st_order_K}
\end{eqnarray}
The corresponding normalization condition is
\begin{eqnarray}
&&\hat{g}_{b}^R(\epsilon_+) \delta \hat{g}^K (\epsilon, \omega) 
+ \delta \hat{g}^K  (\epsilon, \omega) \hat{g}_{b}^A(\epsilon_-) \nonumber \\
&&+ \hat{g}_{b}^K(\epsilon_+) \delta \hat{g}^A (\epsilon, \omega)  +\delta \hat{g}^R (\epsilon, \omega) \hat{g}_{b}^K(\epsilon_-)  =0 . \label{1st_order_norm_K}
\end{eqnarray}
From the gap equation, we derive
\begin{eqnarray}
\delta\hat{\Delta}(\omega) = -\frac{{\mathscr G}}{8}\int d\epsilon  {\rm Tr} [(-i\tau_2) \delta \hat{g}^K(\epsilon,\omega) ]  . \label{gap2}
\end{eqnarray}
The current density induced by the perturbative ac field is given by 
\begin{eqnarray}
&&\delta {\bf J}(\omega) = -i \frac{\sigma_n}{e} \int d\epsilon \delta {\bf S}(\epsilon, \omega) \label{deltaJ}, \\
&& \delta {\bf S} (\epsilon, \omega) = (i/16) {\rm Tr} \bigl[ i{\bf q}_b \times \nonumber \\
&&\bigl\{ \hat{\tau}_3 \hat{g}_b^R(\epsilon_+) \hat{\tau}_3 \delta\hat{g}^K(\epsilon,\omega)
 +\hat{\tau}_3 \delta \hat{g}^R(\epsilon, \omega) \hat{\tau}_3 \hat{g}_b^K(\epsilon_-) \nonumber \\
&& +\hat{\tau}_3 \hat{g}_b^K(\epsilon_+) \hat{\tau}_3 \delta\hat{g}^A(\epsilon,\omega)
 + \hat{\tau}_3 \delta \hat{g}^K(\epsilon, \omega) \hat{\tau}_3 \hat{g}_b^A(\epsilon_-) \} \nonumber \\
&& +i \delta {\bf q}_{\omega} \bigl\{  \hat{\tau}_3 \hat{g}_b^R(\epsilon_+) \hat{\tau}_3 \hat{g}_b^K(\epsilon_-) 
+ \hat{\tau}_3 \hat{g}_b^K(\epsilon_+) \hat{\tau}_3 \hat{g}_b^A(\epsilon_-) 
\bigr\}
\bigr]  , \label{S2} 
\end{eqnarray}
By comparing this with $\delta {\bf J} = \sigma \delta {\bf E}=-i\sigma \hbar \omega \delta {\bf q}_{\omega}/e$, the complex conductivity $\sigma$ can be evaluated.

With this, all the equations are now in place. 
Using Eqs.~(\ref{0thRA})-(\ref{gap2}), we can determine $\hat{g}_b^{R,A,K}$, $\delta \hat{g}^{R,A,K}$, and $\delta \hat{\Delta}$, which allow us to calculate the complex conductivity $\sigma$ via Eqs.~(\ref{deltaJ}) and (\ref{S2}). 
In the following subsections, this process is carried out step by step by expressing all components within the matrices as follows:
$\hat{\Delta} = i\hat{\tau}_2 \Delta$, 
$\hat{g}^R= \hat{\tau}_3 G +  i\hat{\tau}_2 F$, 
and $\hat{g}^A=-\hat{\tau}_3\hat{g}^{R\dagger} \hat{\tau}_3$. 
Section~\ref{section_0th_order} solves the zeroth-order equations to obtain $\hat{g}_b^{R,A,K}$. Section~\ref{section_1st_order} then addresses the first-order equations to derive $\delta \hat{g}^{R,A,K}$ and $\delta \hat{\Delta}$.
In Section~\ref{section_sigma}, we derive the complex conductivity formula.

%%%%%%%%%%%%%%%%%%%
\subsection{Equilibrium Green's Functions in the Presence of an arbitrary strength of bias dc} \label{section_0th_order}
%%%%%%%%%%%%%%%%%%%

The zeroth-order equation [Eq.~(\ref{0thRA})] simplifies to the well-known dc-carrying Usadel equation:  
$\Delta_b G_b - \epsilon F_b = is G_b F_b$.  
By numerically solving this equation for a given $s$, we construct a table of $G_b(\epsilon)$ and $F_b(\epsilon)$ for all relevant $\epsilon$ values. These results are essential for calculating the nonequilibrium Green's functions and the Higgs mode in Section~\ref{section_1st_order}, as well as the complex conductivity in Section~\ref{section_sigma}. 
Although this equation has been solved in numerous studies over the past decades, we provide a brief explanation of our solution method here for the reader's convenience.

We begin by converting the Usadel equation into the Matsubara representation:  
\begin{eqnarray}
\Delta_b G_m - \hbar \omega_m F_m = s G_m F_m, \label{eqUsadel2}
\end{eqnarray}
where $G_m = G_b(i\hbar \omega_m)$, $F_m = i F_b(i\hbar \omega_m)$, and $\hbar \omega_m = 2\pi kT (m+1/2)$ is the Matsubara frequency. The normalization condition in the Matsubara representation becomes $G_m^2 + F_m^2 = 1$.
The order parameter $\Delta_b$ under a bias dc is determined by the gap equation [Eq.~(\ref{gap_bias})], which can also be expressed as:  
\begin{eqnarray}
\ln \frac{T_{c0}}{T} = 2\pi kT \sum_{\omega_m > 0} \Bigl( \frac{1}{\hbar \omega_m} - \frac{F_m}{\Delta_b} \Bigr). \label{gap_bias2}
\end{eqnarray}
Solving Eqs.~(\ref{eqUsadel2}) and (\ref{gap_bias2}) to determine $\Delta_b$, $G_m$, and $F_m$ is a straightforward task. Once these quantities are obtained, the bias dc current can be calculated using the following relations:  
\begin{eqnarray}
&&\frac{J_b(s, T)}{J_0} = \sqrt{\pi \frac{s}{\Delta_0}} \frac{n_s(s, T)}{n_{s0}}  , \label{Jb} \\
&&\frac{n_s(s, T)}{n_{s0}} = \frac{\lambda_0^2}{\lambda^2(s, T)} = \frac{4kT}{\Delta_0} \sum_{\omega_m > 0} F_m^2 . \label{rho_s} 
\end{eqnarray}
Here, $J_0=H_{c0}/\lambda_0$, $n_{s0} = 2\pi m N_0 D \Delta_0 / \hbar$, $\lambda_0 = \lambda(0, 0) = \sqrt{\hbar / \pi \mu_0 \Delta_0 \sigma_n}$, and $H_{c0} = \sqrt{N_0 / \mu_0} \Delta_0$. 
Note that Eq.~(\ref{Jb}) provides the relationship between the bias superflow parameter $s$ and the bias current $J_b$, which is essential for translating results expressed as a function of $s$ into those expressed as a function of $J_b$ (See also Appendix~\ref{appendix_1}).

\begin{figure}[tb]
   \begin{center}
   \includegraphics[height=0.45\linewidth]{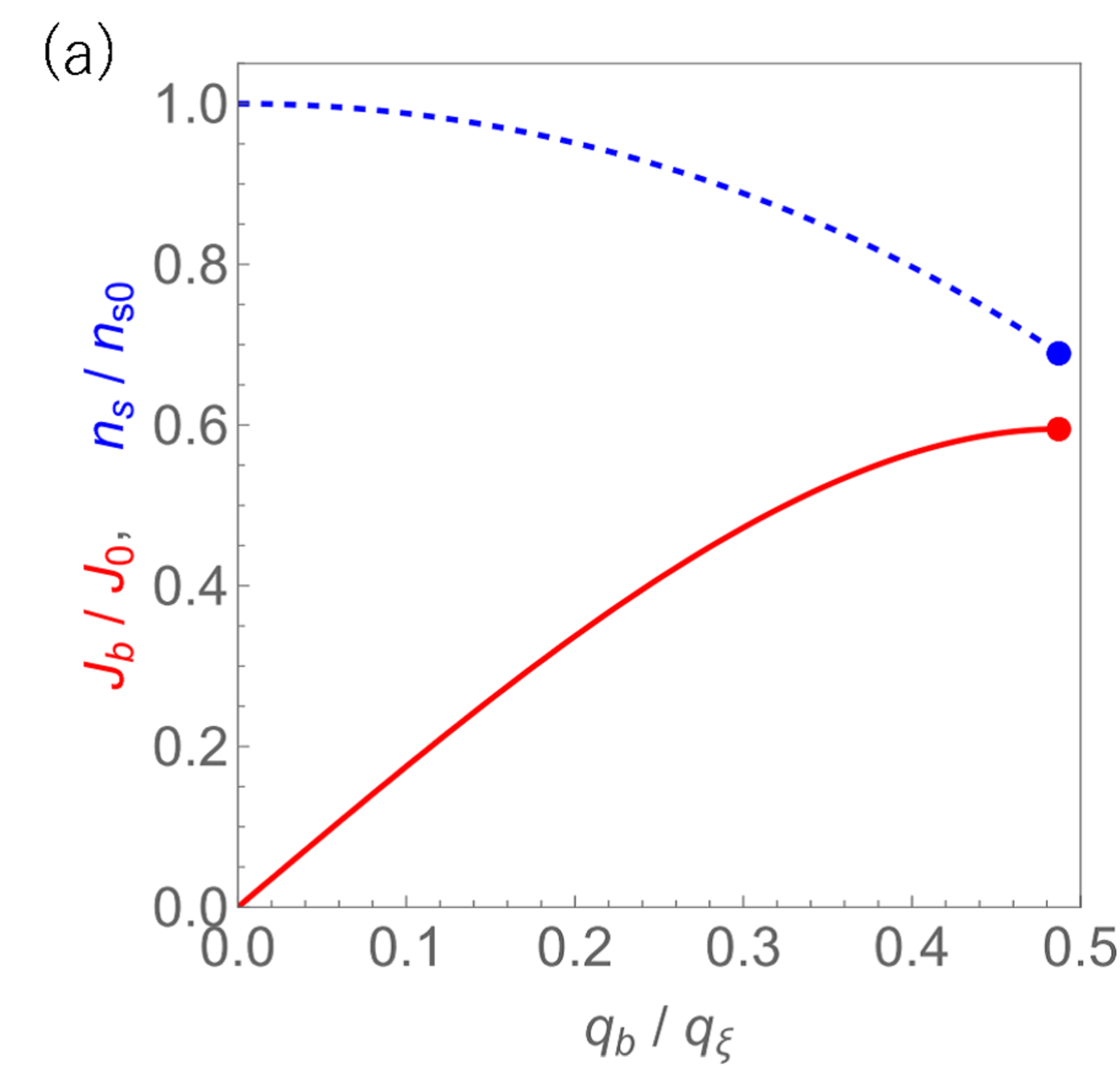}
   \includegraphics[height=0.45\linewidth]{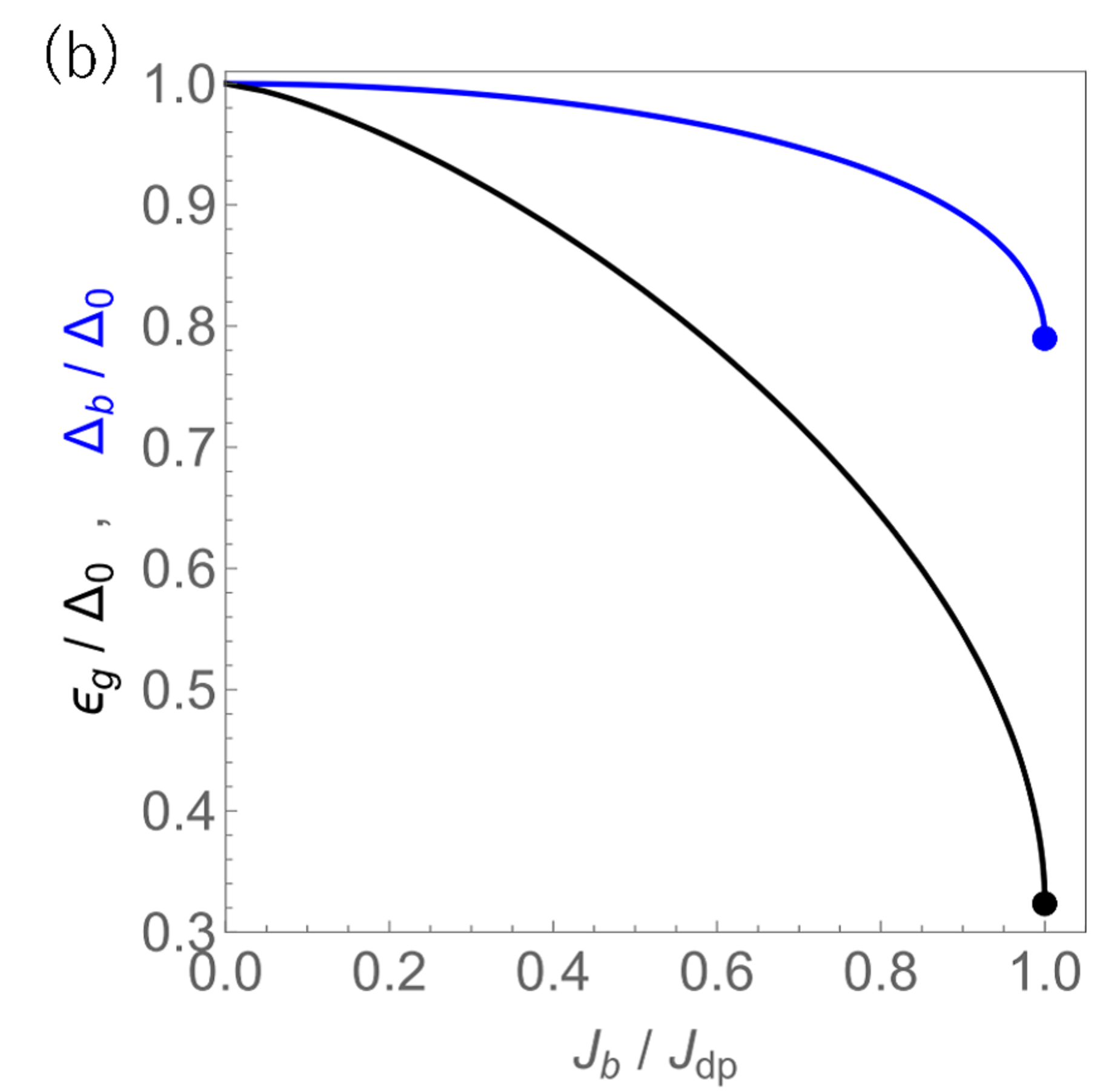}
   \includegraphics[height=0.45\linewidth]{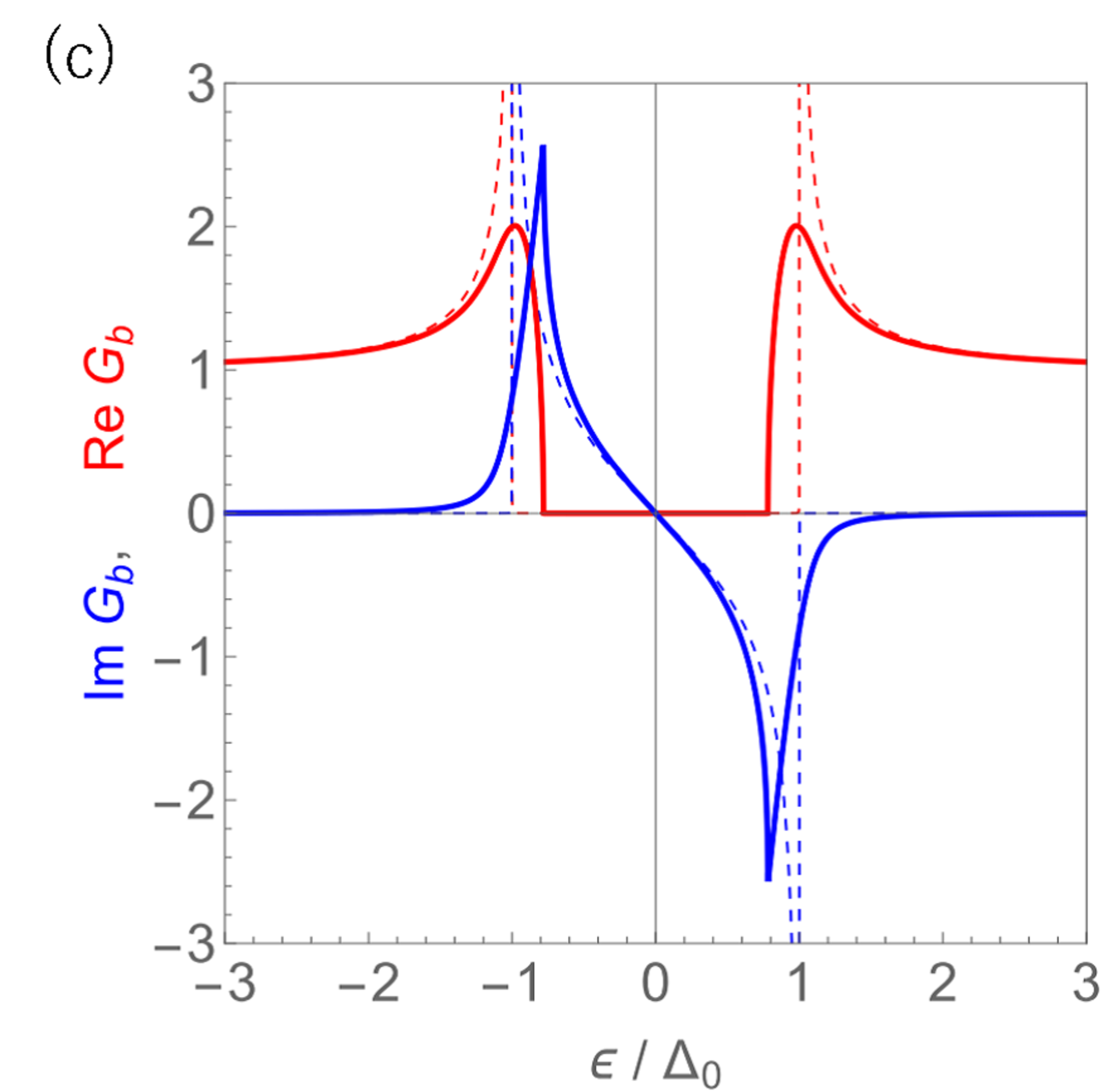}
   \includegraphics[height=0.45\linewidth]{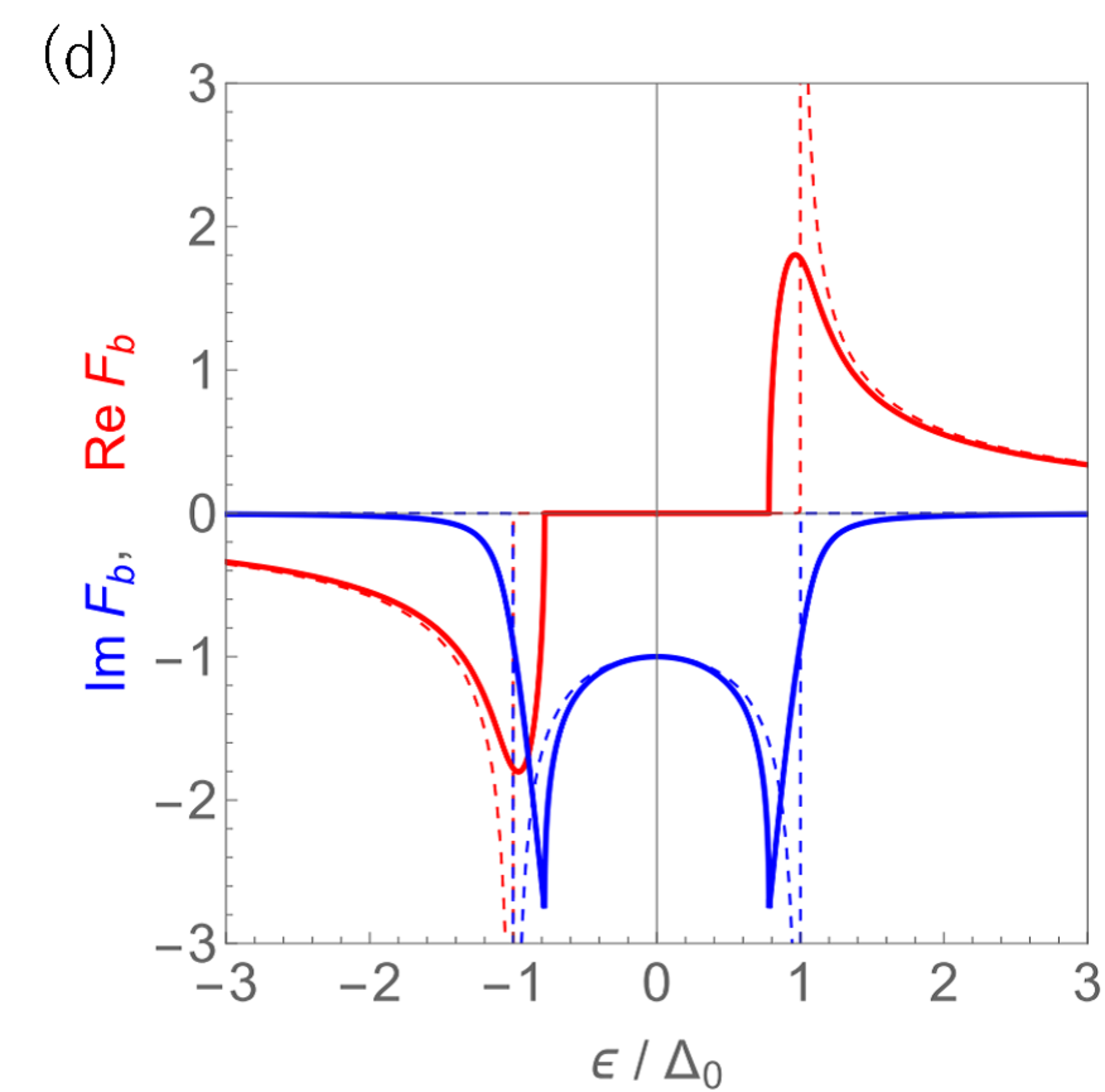}
   \end{center}\vspace{0 cm}
   \caption{
Equilibrium properties of a superconducting film under a bias dc at zero temperature.  
(a) Bias dc current density $J_b$ as a function of the bias superfluid momentum $q_b$. The dot indicates the depairing current density $J_{\rm dp}$.  
(b) Pair potential $\Delta_b$ (blue) and spectral gap $\epsilon_g$ (black) as functions of the bias dc current density $J_b$. The dot indicates the equilibrium depairing current density $J_{\rm dp}$.
(c) Real and imaginary parts of $G_b$ for $J_b = 0$ (dashed) and $J_b = 0.6 J_{\rm dp}$ (solid).  
(d) Real and imaginary parts of $F_b$ for $J_b = 0$ (dashed) and $J_b = 0.6 J_{\rm dp}$ (solid).  
   }\label{fig2}
\end{figure}

Figure~\ref{fig2}(a) depicts the bias dc ($J_b$) as a function of the normalized bias momentum $q_b/q_{\xi} = \sqrt{s/\Delta_0}$, providing a means to translate momentum into current.  
The dot indicates the well-known dirty-limit equilibrium depairing current density, $J_{\rm dp}(0) = 0.595 H_{c0} / \lambda_0$, derived decades ago~\cite{Maki, Maki_book, KL} (See also Appendix~\ref{appendix_1}).  
Figure~\ref{fig2}(b) shows the pair potential $\Delta_b$ under a bias dc as a function of $J_b$, up to the depairing current density (blue curve).

Having determined $\Delta_b$ for a given $s$ through the Matsubara representation calculations described above, solving the real-frequency representation of the Usadel equation to obtain the equilibrium Green's functions under a bias dc ($G_b$ and $F_b$) becomes straightforward. 
First, the spectral gap, defined as the maximum energy $\epsilon$ for which the quasiparticle density of states vanishes, is given by~\cite{Maki, Maki_book}
\begin{eqnarray}
\epsilon_g = \Bigl( \Delta_b^{\frac{2}{3}} - s^{\frac{2}{3}} \Bigr)^{\frac{3}{2}} ,
\end{eqnarray}
and is illustrated in Fig.~\ref{fig2}(b) (black curve). 
To calculate the spectrum, we rewrite the real-frequency representation of the Usadel equation using the parametrization 
\( G_b = \cosh(u + i v) \) and \( F_b = \sinh(u + i v) \), which transforms the problem into two equations for \( u \) and \( v \):  
\begin{eqnarray}
&& s  \left( \epsilon \cosh u + \Delta_b \sinh u \right) \cos^3 v 
+ \Gamma \epsilon \cos^2 v \nonumber  \\
&&- s \left( \epsilon \cosh^3 u - \Delta_b \sinh^3 u \right) \cos v \nonumber \\
&&- \Gamma \cosh u \cdot \left( \epsilon \cosh u - \Delta_b \sinh u \right)=0, \label{real_freq_usadel_1}\\
&& s \cos v \cosh^3 u + \Gamma \cosh^2 u - s \cos^3 v \cosh u \nonumber \\
&&+ \left( \Delta_b \sin v - \Gamma \cos v \right) \cos v=0 . \label{real_freq_usadel_2}
\end{eqnarray}
Here, \( \Gamma \) represents a small damping factor (e.g., \( \Gamma / \Delta_0 = 10^{-5} \)), introduced to stabilize the numerical computation and facilitate convergence.

The solutions to Eqs.~(\ref{real_freq_usadel_1}) and (\ref{real_freq_usadel_2}) can be obtained in a straightforward manner. Representative examples of \( G_b \) and \( F_b \) calculated using this method are shown in Fig.~\ref{fig2}(c) and Fig.~\ref{fig2}(d), respectively. 
Furthermore, it is easy to confirm that the spectral gap $\epsilon_g$ determined in Fig.~\ref{fig2}(b) aligns consistently with the gap edges presented in Fig.~\ref{fig2}(c).

At this stage, we can determine $G_b(\epsilon)$ and $F_b(\epsilon)$ for a given $s$ or $J_b$. 
The next step is to express the nonequilibrium Green's functions in terms of $G_b(\epsilon)$ and $F_b(\epsilon)$.

%%%%%%%%%%%%%%%%%%%
\subsection{Nonequilibrium Green's functions and the Higgs mode} \label{section_1st_order}
%%%%%%%%%%%%%%%%%%%

Our next task is to solve the first-order equations for the R (A) and K components [Eqs.~(\ref{1st_order_RA}) and (\ref{1st_order_K})]. Although the calculations are lengthy, they are straightforward and yield relatively simple solutions expressed in terms of $G_b$ and $F_b$.

The $R (A)$ component's first-order equation [Eq.~(\ref{1st_order_RA})] and the normalization condition [Eq.~(\ref{1st_order_norm_RA})] lead to:
\begin{eqnarray}
&& \delta G (\epsilon, \omega) = \frac{F_{b+} + F_{b-} }{G_{b+} + G_{b-}} \delta F (\epsilon, \omega), \label{deltaG} \\
&& \delta F (\epsilon, \omega) = \zeta \delta \Delta(\omega) + \kappa \delta W , \label{dF}\\
&& \zeta = -\frac{G_{b+}  + G_{b-}}{\hbar \omega}\frac{F_{b+} -F_{b-}}{F_{b+} + F_{b-}} , \label{zeta} \\
&& \kappa =  i \frac{G_{b+} + G_{b-}}{\hbar \omega} ( F_{b+} - F_{b-} ) , \label{kappa}
\end{eqnarray}
where $G_{b\pm} = G_b(\epsilon \pm \hbar \omega / 2)$ and $F_{b\pm} = F_b(\epsilon \pm \hbar \omega / 2)$.

The $K$ component's first-order equation [Eq.~(\ref{1st_order_K})] and the normalization condition [Eq.~(\ref{1st_order_norm_K})] can be solved by expressing $\delta \hat{g}^K$ in terms of the {\it anomalous} term $\delta \hat{g}^a$, following Eliashberg's approach~\cite{Gorkov, Eliashberg}:  
\begin{eqnarray}
\delta \hat{g}^K = \delta \hat{g}^R \mathcal{T}_- - \delta \hat{g}^A \mathcal{T}_+ + \delta \hat{g}^a (\mathcal{T}_+ - \mathcal{T}_-).
\end{eqnarray}
Substituting this form and performing a lengthy calculation, we obtain
\begin{eqnarray}
&& \delta G^a (\epsilon, \omega) = \frac{F_{b+} - F_{b-}^*}{G_{b+} - G_{b-}^*} \delta F^a (\epsilon, \omega), \\
&& \delta F^a (\epsilon, \omega) =  \zeta^a \delta\Delta + \kappa^a \delta W, \label{dFa}\\
&& \zeta^a = -\frac{G_{b+} - G_{b-}^*}{\hbar \omega} \frac{F_{b+}  + F_{b-}^*}{F_{b+} - F_{b-}^*}, \label{zeta_ano}\\
&& \kappa^a = i\frac{G_{b+} - G_{b-}^*}{\hbar \omega} (F_{b+} + F_{b-}^*) .\label{kappa_ano}
\end{eqnarray}
With $\delta \hat{g}^K$ now expressed in terms of $G_b$, $F_b$, $\delta W$, and $\delta \Delta$, we substitute it into the perturbative part of the gap equation [Eq.~(\ref{gap2})]. 
This yields:  
\begin{eqnarray}
\delta \Delta (\omega) &=& \Psi \delta W,  \label{Higgs_response} \\
\Psi &=& \frac{-({\mathscr G}/4) \int d\epsilon \psi_N(\epsilon)}{1+({\mathscr G}/4)\int d\epsilon \psi_D(\epsilon)}, \label{Psi}\\
\psi_N &=& \kappa_{(\Psi)} \mathcal{T}_-  + \kappa_{(\Psi)}^*\mathcal{T}_+  
 + ( \mathcal{T}_+ - \mathcal{T}_- ) \kappa_{(\Psi)}^a , \\
\psi_D &=& \zeta \mathcal{T}_-  + \zeta^* \mathcal{T}_+ + ( \mathcal{T}_+ - \mathcal{T}_- )\zeta^a , \label{psiD}
\end{eqnarray}
Note that in the \({\rm ac} \perp {\rm dc}\) (i.e., \(\delta {\bf q} \perp {\bf q}_b\)) configuration, $\delta W \propto {\bf q}_b \cdot \delta {\bf q}_{\omega}$ vanishes. As a result, all nonequilibrium corrections, including $\delta G$, $\delta F$, $\delta G^a$, $\delta F^a$, and $\delta \Delta$, are eliminated.

At this stage, we have solved the Keldysh-Usadel equation and expressed the solution in terms of \( G_b \) and \( F_b \) for an arbitrary strength of the bias dc.  
A crucial sanity check, performed by taking the weak-bias dc limit, verifies that our results are fully consistent with the earlier study by Moor et al.~\cite{Moor}, where both the ac and dc fields were treated as perturbations (see Appendix~\ref{appendix_2}).  
Furthermore, in the short mean free path limit, our results correspond to those obtained in the more general framework of Refs.~\cite{Jujo, 2024_Kubo}, which considered arbitrary mean free paths.  

The next step is to derive an explicit expression for the complex conductivity formula based on these solutions.

%%%%%%%%%%%%%%%%%%%
\subsection{Complex conductivity under a weak ac field superposed on an arbitrary strength of bias dc} \label{section_sigma}
%%%%%%%%%%%%%%%%%%%

We examine two specific configurations: one where the ac perturbation is parallel to the dc bias (\({\rm ac} \parallel {\rm dc}\), i.e., \(\delta {\bf q} \parallel {\bf q}_b\)), and the other where it is perpendicular (\({\rm ac} \perp {\rm dc}\), i.e., \(\delta {\bf q} \perp {\bf q}_b\)).
By substituting the nonequilibrium Green's functions [Eqs.~(\ref{deltaG})-(\ref{psiD})] into the expression for the current response, we arrive at one of the main results of this paper: $\sigma = (\sigma_n/\hbar\omega) \int d\epsilon (\delta S/\delta q_{\omega})$ or 
\begin{eqnarray}
\sigma 
=
\begin{cases}
\sigma^{(0)} + \sigma^{(1)} + \sigma^{(2)}  & ({\rm ac} \parallel {\rm dc}) , \\
\sigma^{(0)}  & ({\rm ac} \perp {\rm dc}) , 
\end{cases} \label{total_sigma}
\end{eqnarray}
where
\begin{eqnarray}
&&\frac{\sigma^{(0)}}{\sigma_n} = \int\!\! \frac{d\epsilon}{\hbar\omega} ({\rm Re}G_b \, {\rm Re}G_b' + {\rm Re}F_b \, {\rm Re}F_b' )  
(f_{\rm FD} - f_{\rm FD}' ) \nonumber \\ 
&& + i \int\!\! \frac{d\epsilon}{\hbar\omega} ({\rm Re}G_b \, {\rm Im}G_b' + {\rm Re}F_b \,  {\rm Im}F_b' ) (2f_{\rm FD} -1 ) , \label{sigma0} \\
&& \frac{\sigma^{(1)}}{\sigma_n} = \frac{8s}{\hbar\omega} \int\!\! \frac{d\epsilon}{\hbar\omega}
 {\rm Re} F_b \, {\rm Im} F_b \, {\rm Re} G_{b}' ( f_{\rm FD} - f_{\rm FD}' ) \nonumber \\
&&+ i  \frac{2s}{\hbar\omega} \int\!\! \frac{d\epsilon}{\hbar\omega} \Bigl[ 
2 {\rm Re}F_b\,  {\rm Im}F_b\, {\rm Im} \bigl\{ G_b + G_{b}' \bigr\} 
 + \bigl\{ ({\rm Re} F_b')^{2}  \nonumber \\
&&-({\rm Re} F_b)^2 + ({\rm Im} F_b)^2 - ({\rm Im} F_b')^2 \bigr\} {\rm Re}G_b \Bigr]  (2f_{\rm FD} -1) , \label{sigma1} \\
&& \frac{\sigma^{(2)}}{\sigma_n} = \frac{2 s\Psi}{\hbar\omega} \int\!\! \frac{d\epsilon}{\hbar\omega} ( {\rm Re}F_b \, {\rm Re}G_b' - {\rm Re}G_b \, {\rm Re}F_b'  ) 
( f_{\rm FD} - f_{\rm FD}' ) \nonumber \\
&&+ i \frac{2s\Psi}{\hbar\omega} \int\!\! \frac{d\epsilon}{\hbar\omega}  \bigl\{ {\rm Re}G_b \, {\rm Im}(F_b - F_b' ) +  {\rm Re}F_b \, {\rm Im}(G_b + G_b' ) \bigr\} \nonumber \\
&& \times ( 2f_{\rm FD}-1 ) . \label{sigma2}
\end{eqnarray}
Here, $G_b'$, $F_b'$, and $f_{\rm FD}'$ denote $G_b(\epsilon + \hbar \omega)$, $F_b(\epsilon + \hbar \omega)$, and $f_{\rm FD}(\epsilon + \hbar \omega)$, respectively.  
The functions $G_b$ and $F_b$ are the equilibrium Green's functions under a bias dc, calculated using Eqs.~(\ref{real_freq_usadel_1}) and (\ref{real_freq_usadel_2}). 
This formula represents the dirty-limit counterpart of the previously derived expression~\cite{Jujo, 2024_Kubo} applicable to arbitrary mean free paths using the Keldysh-Eilenberger theory and is significantly simpler by comparison.

It is worth noting that $\sigma^{(0)}$ closely resembles the well-known formula for the complex conductivity in the absence of a bias dc~\cite{Nam, Gurevich_Kubo, 2022_Kubo}.  
This term represents a straightforward extension of the zero-bias dc case to the finite-bias dc case, incorporating $G_b$ and $F_b$, which are the equilibrium Green's functions under a bias dc, instead of the zero-current Green's functions.  
However, a rigorous derivation based on the Keldysh-Usadel equation reveals that this naive extension is insufficient, and additional contributions, $\sigma^{(1)}$ and $\sigma^{(2)}$ [Eqs.~(\ref{sigma1}) and (\ref{sigma2}), respectively] are required for the ${\rm ac} \parallel {\rm dc}$ case.

%%%%%%%%%%%%%%%%%%%
%%%%%%%%%%%%%%%%%%%
\section{Higgs mode and complex conductivity under a bias dc}\label{section_Higgs_and_sigma}
%%%%%%%%%%%%%%%%%%%
%%%%%%%%%%%%%%%%%%%

%%%%%%%%%%%%%%%%%%%
\subsection{Sanity Check: Effect of Perturbative Bias dc} \label{weak_dc_sigma}
%%%%%%%%%%%%%%%%%%%

\begin{figure}[tb]
   \begin{center}
   \includegraphics[height=0.46\linewidth]{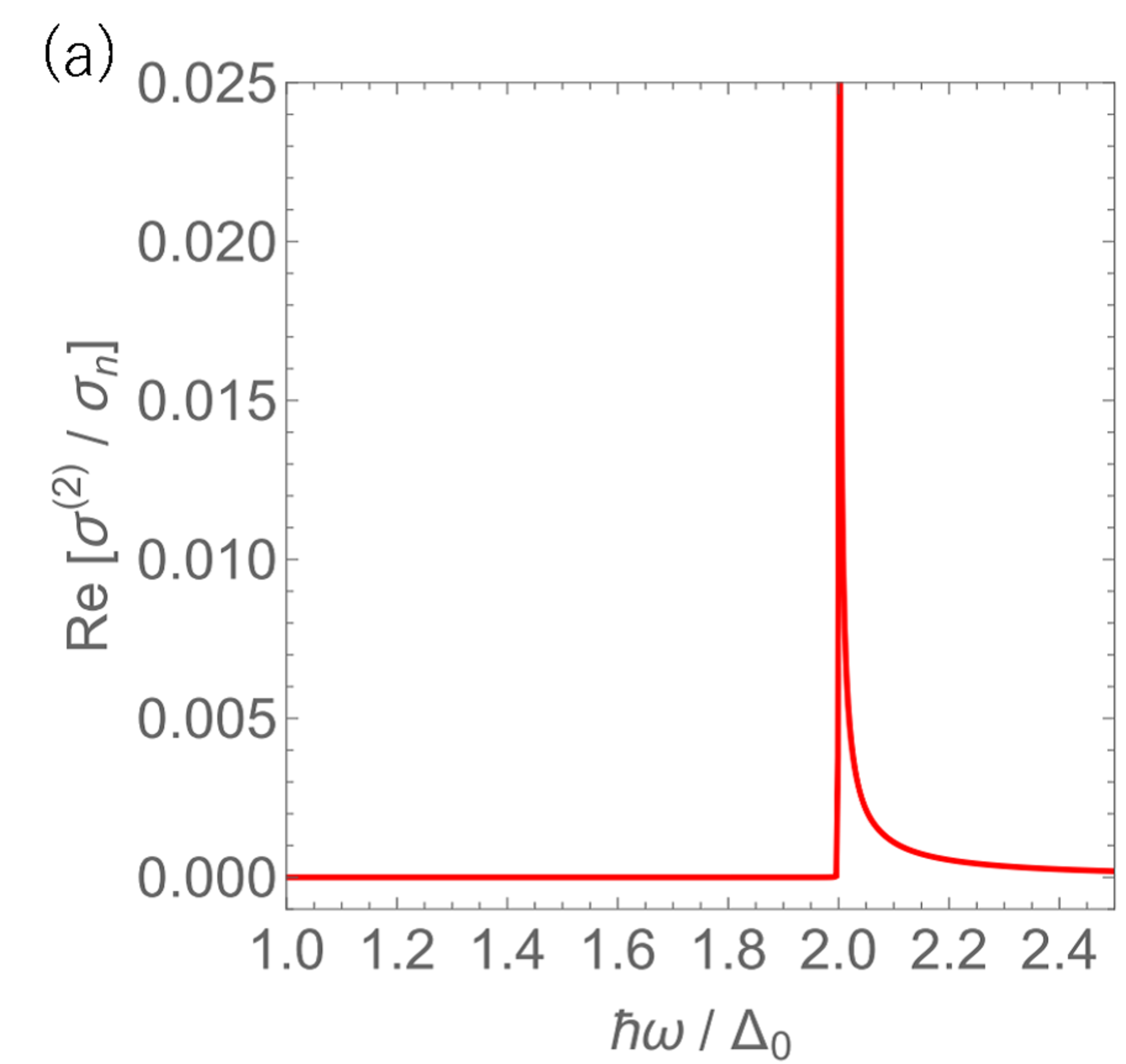}
   \includegraphics[height=0.46\linewidth]{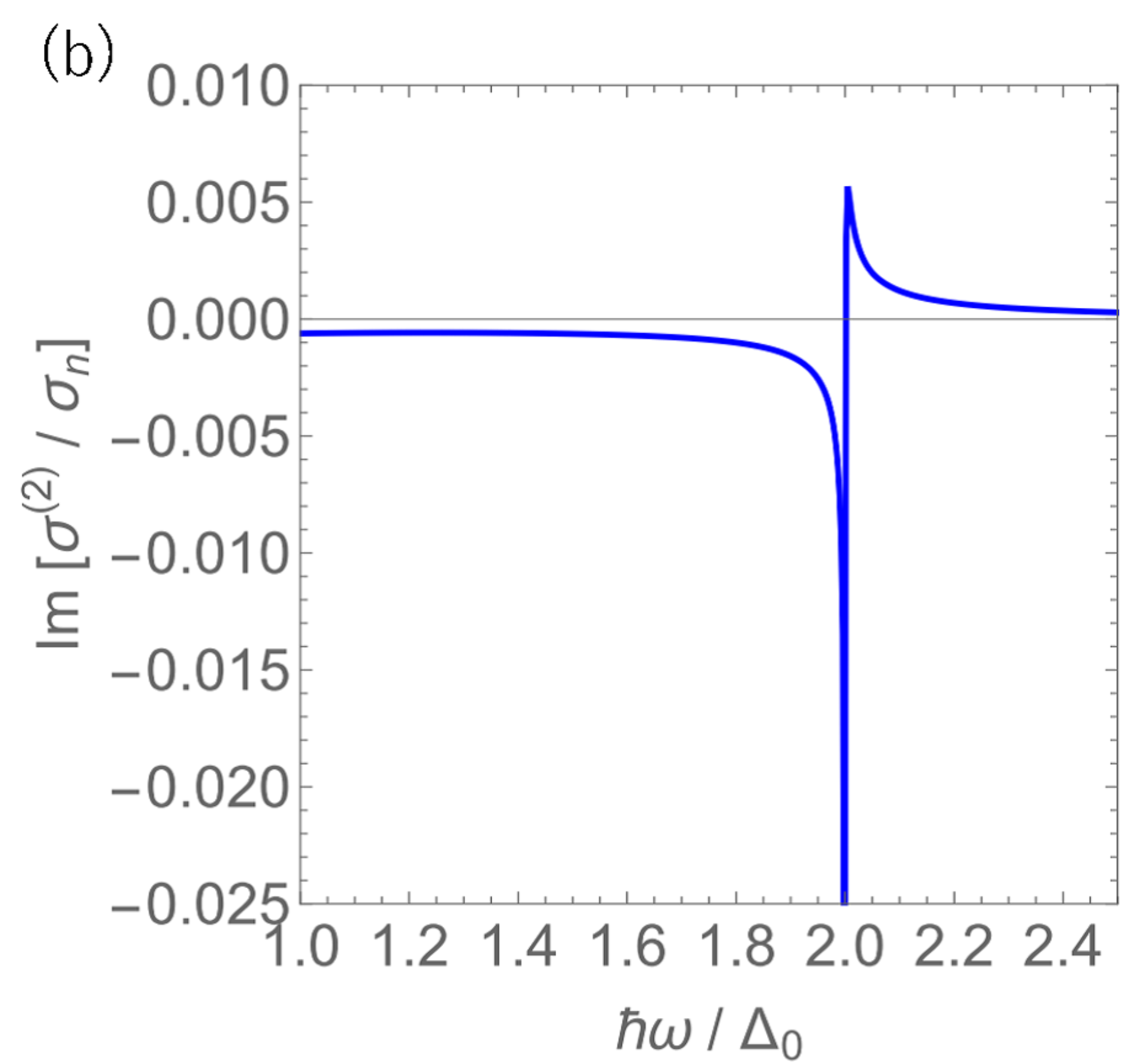}
   \end{center}\vspace{0 cm}
   \caption{
Sanity check: manifestation of the Higgs mode in the complex conductivity under a perturbative dc bias.  
(a) Real part and (b) imaginary part of $\sigma^{(2)}$ calculated for ${\rm ac} \parallel {\rm dc}$, $T=0$, and $J_b/J_{\rm dp} \sim q_b / q_{\xi} =\sqrt{s / \Delta_0}= 0.01$.  
   }\label{fig3}
\end{figure}

As discussed in Section~\ref{intro}, when an ac field is superposed parallel to a bias dc (${\rm ac} \parallel {\rm dc}$), the Higgs mode responds linearly to the ac field [see Fig.~\ref{fig1}(a) and Eq.~(\ref{Higgs_response})], with the bias dc serving as a tuning knob to amplify the Higgs mode response.  
The most straightforward way to examine the Higgs mode contribution is to consider the perturbative bias dc case and calculate ${\rm Re} \sigma^{(2)}$ and ${\rm Im} \sigma^{(2)}$ to leading order, $\mathcal{O}(s)$. In this regime, the zero-current Green's functions, $G_0 = (\epsilon + i0) / \sqrt{(\epsilon + i0)^2 - \Delta^2}$ and $F_0 = \Delta / \sqrt{(\epsilon + i0)^2 - \Delta^2}$, can be substituted for $G_b$ and $F_b$, respectively.  
This simplified case was studied in the pioneering work by Moor et al.~\cite{Moor}, who investigated the Higgs mode in dc-biased dirty superconductors, treating both the ac field and the dc bias as perturbations.
Figure~\ref{fig3} shows the calculated results for $\sigma^{(2)}$, where the Higgs mode manifests as a characteristic peak and dip at $\hbar \omega = 2\Delta_0$ in ${\rm Re} \sigma^{(2)}$ and ${\rm Im} \sigma^{(2)}$, respectively.

%%%%%%%%%%%%%%%%%%%
\subsection{Nonperturbative Effects of a Strong Bias dc and Higgs-Induced Instability}\label{strong_dc_sigma}
%%%%%%%%%%%%%%%%%%%

\begin{figure}[tb]
   \begin{center}
   \includegraphics[height=0.46\linewidth]{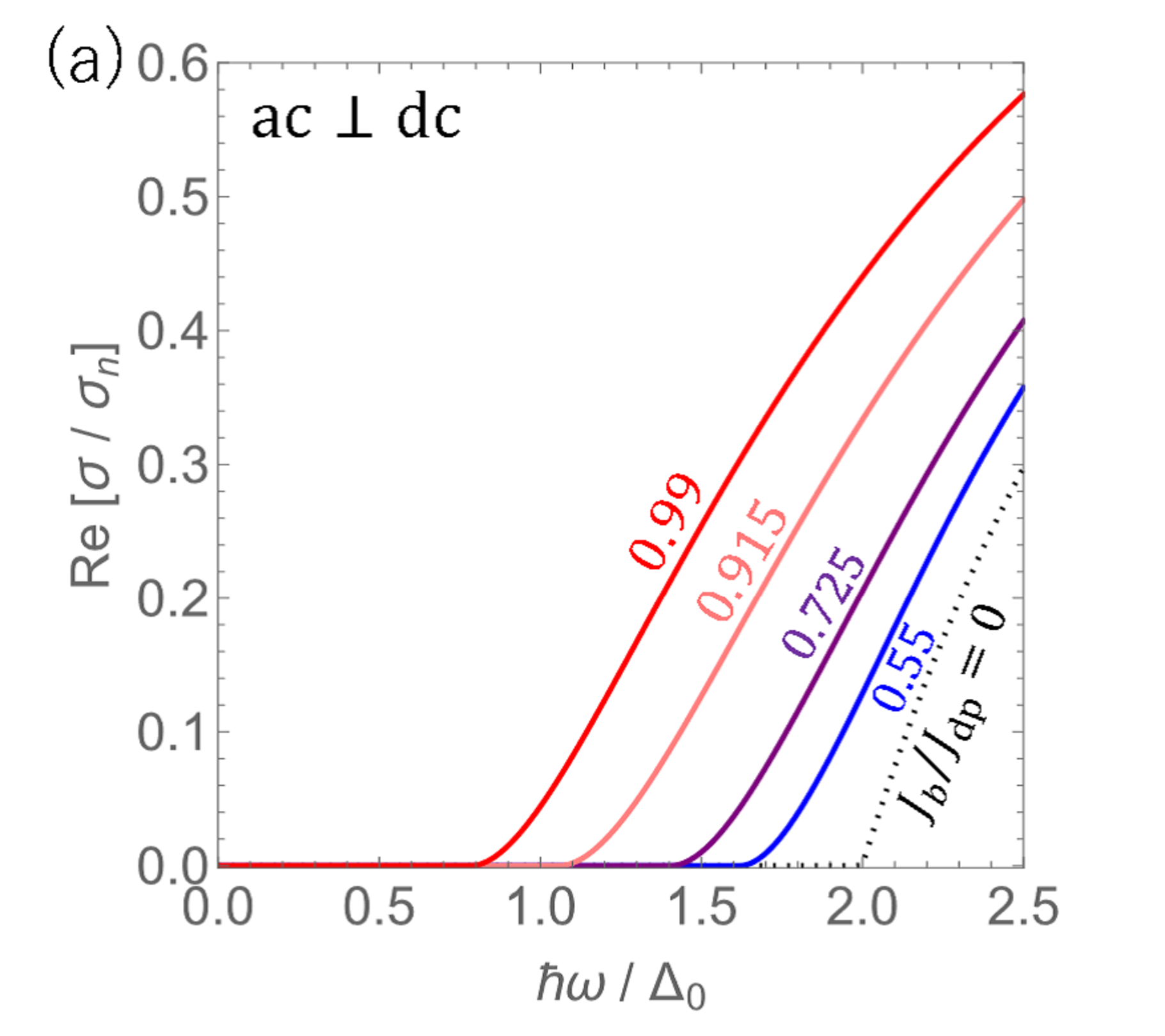}
   \includegraphics[height=0.46\linewidth]{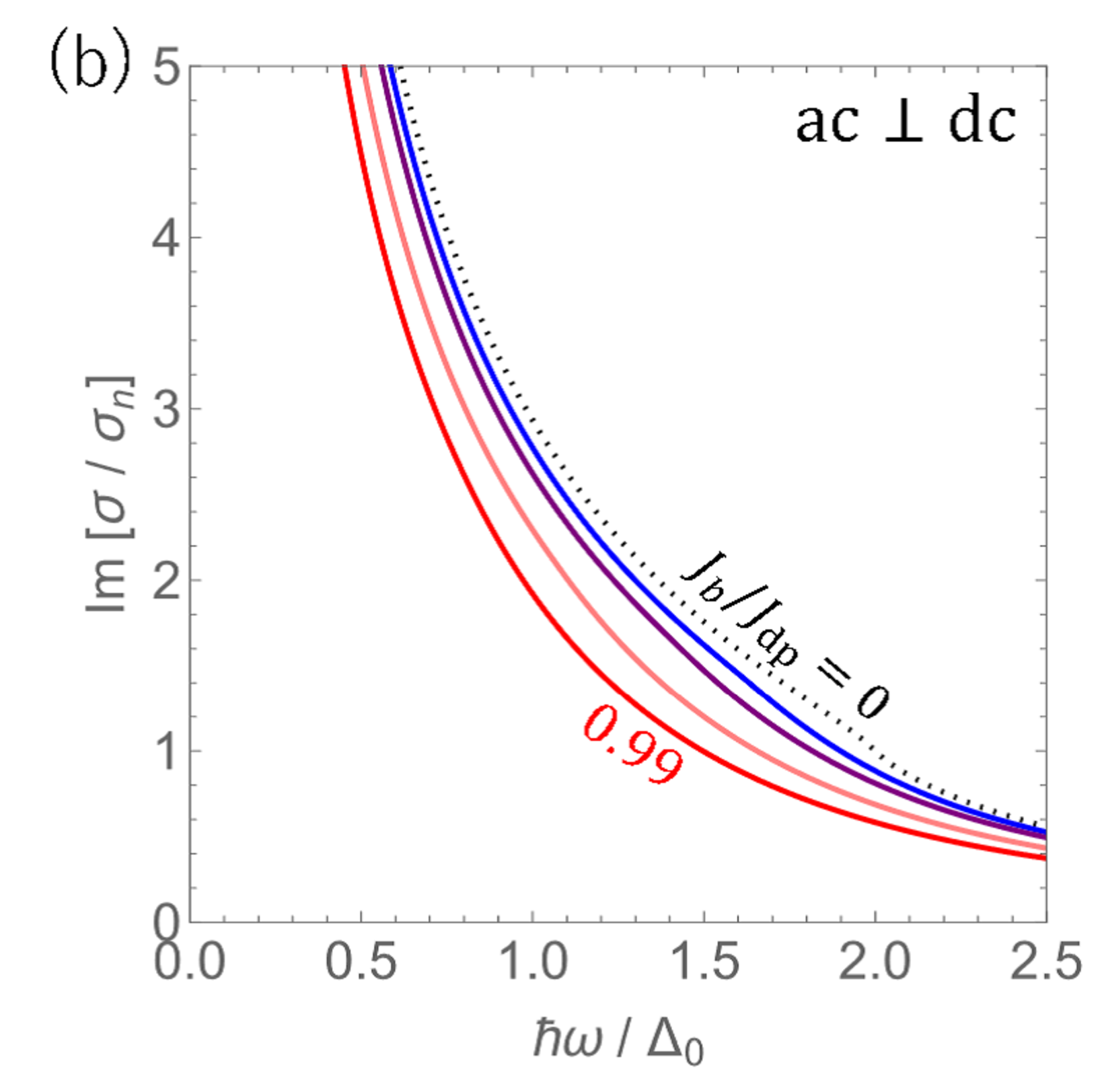}
   \end{center}\vspace{0 cm}
   \caption{
Absence of Higgs resonance in the complex conductivity for the \( {\rm ac} \perp {\rm dc} \) configuration.  
(a) Real part and (b) imaginary part of the complex conductivity calculated at \( T = 0 \). 
   }\label{fig4}
\end{figure}

Now, we explore the impact of stronger bias dc on the complex conductivity, leveraging our previously derived formula [Eqs.~(\ref{sigma0})-(\ref{sigma2})], which remains valid for arbitrary bias dc strengths, including those approaching the depairing current.
The calculation procedure is straightforward: numerically compute the equilibrium Green's functions under a bias dc ($G_b$ and $F_b$) for a given $s$ or $J_b$ using the method outlined in Sec.~\ref{section_0th_order}, and substitute them into Eqs.~(\ref{sigma0})-(\ref{sigma2}).

To begin with, Figure~\ref{fig4} presents the complex conductivity for the \( {\rm ac} \perp {\rm dc} \) configuration, calculated at \( T=0 \).  
In this configuration, the response is solely determined by \( \sigma^{(0)} \).  
As a result, the complex conductivity exhibits a monotonic dependence on both frequency and dc bias strength, with no Higgs resonance features.  
Additionally, the onset of \( {\rm Re} \, \sigma \) in Figure~\ref{fig4}(a) is given by \( \hbar \omega = 2\epsilon_g \), where \( \epsilon_g(J_b) \) is shown in Figure~\ref{fig2}(b).

\begin{figure*}[bth]
   \begin{center}
   \includegraphics[height=0.23\linewidth]{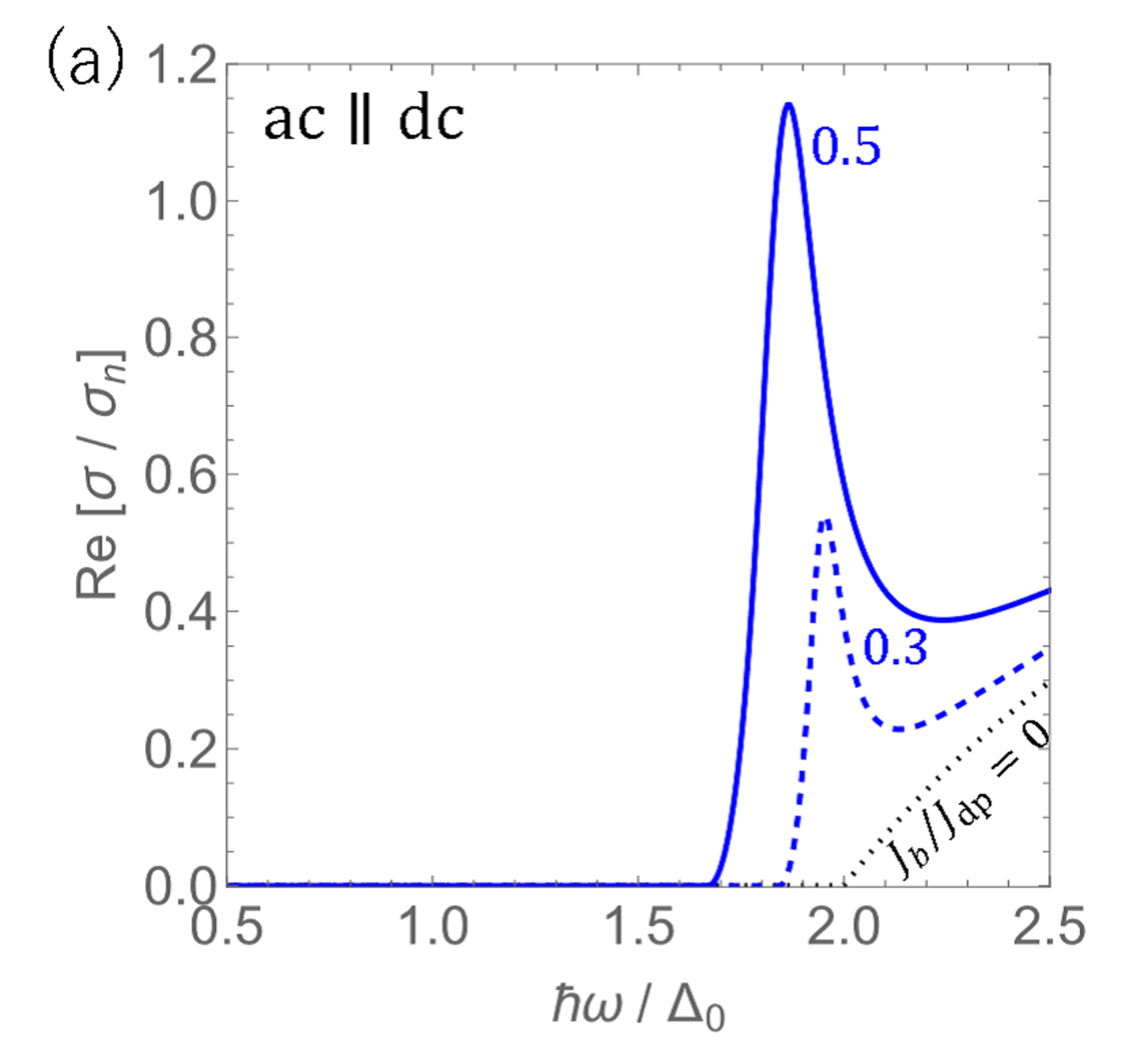}
   \includegraphics[height=0.24\linewidth]{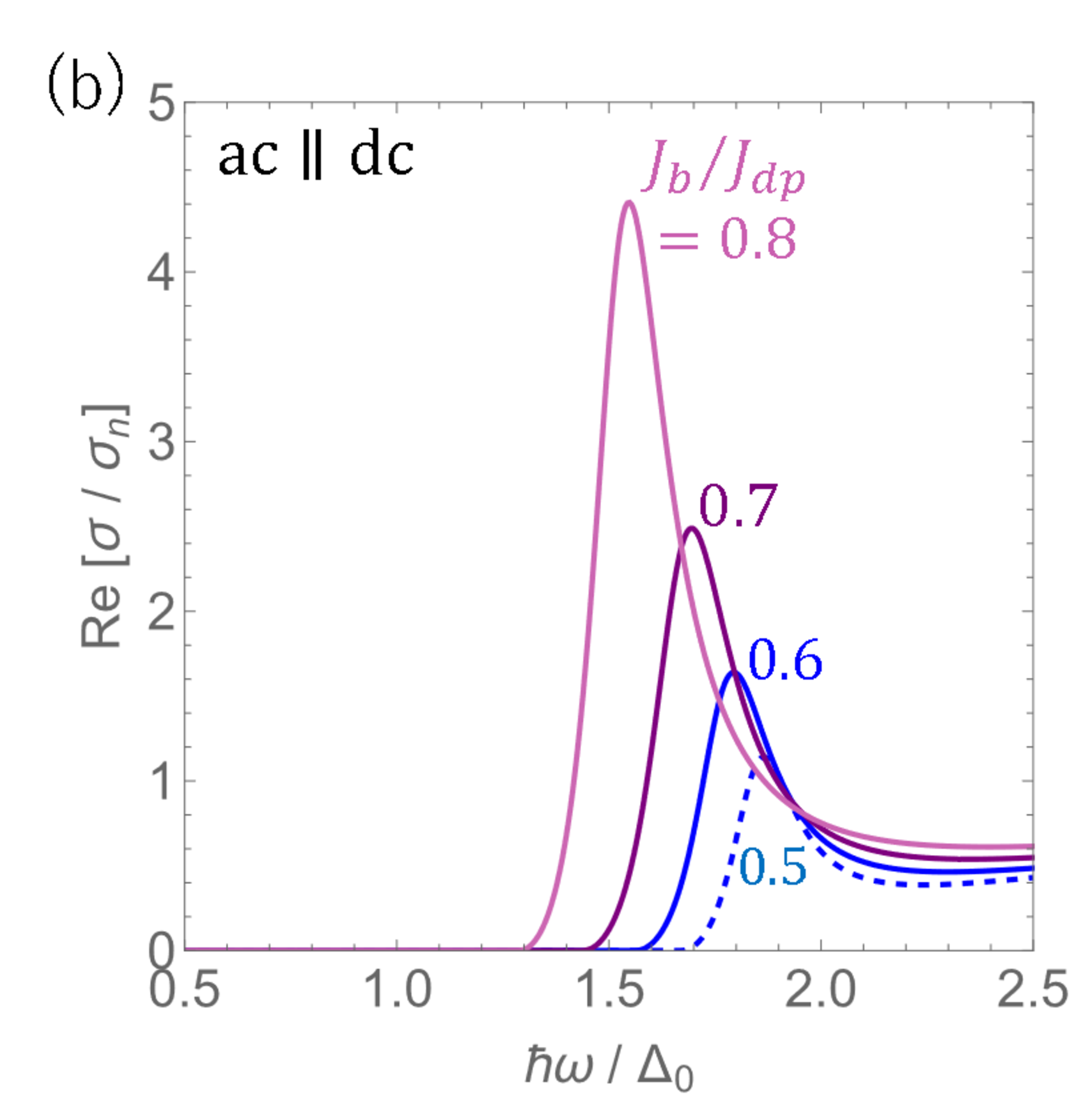}
   \includegraphics[height=0.23\linewidth]{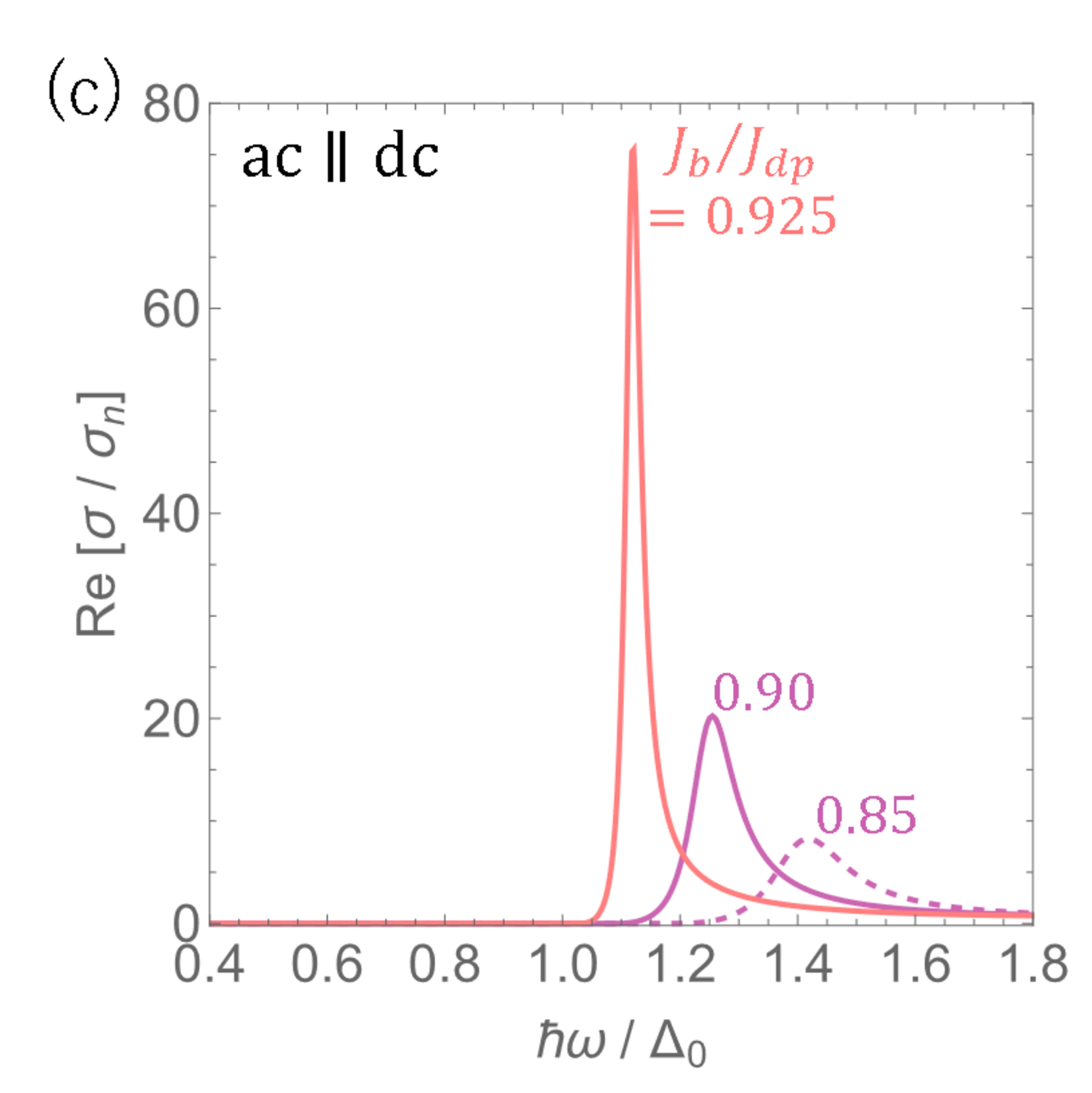}
   \includegraphics[height=0.23\linewidth]{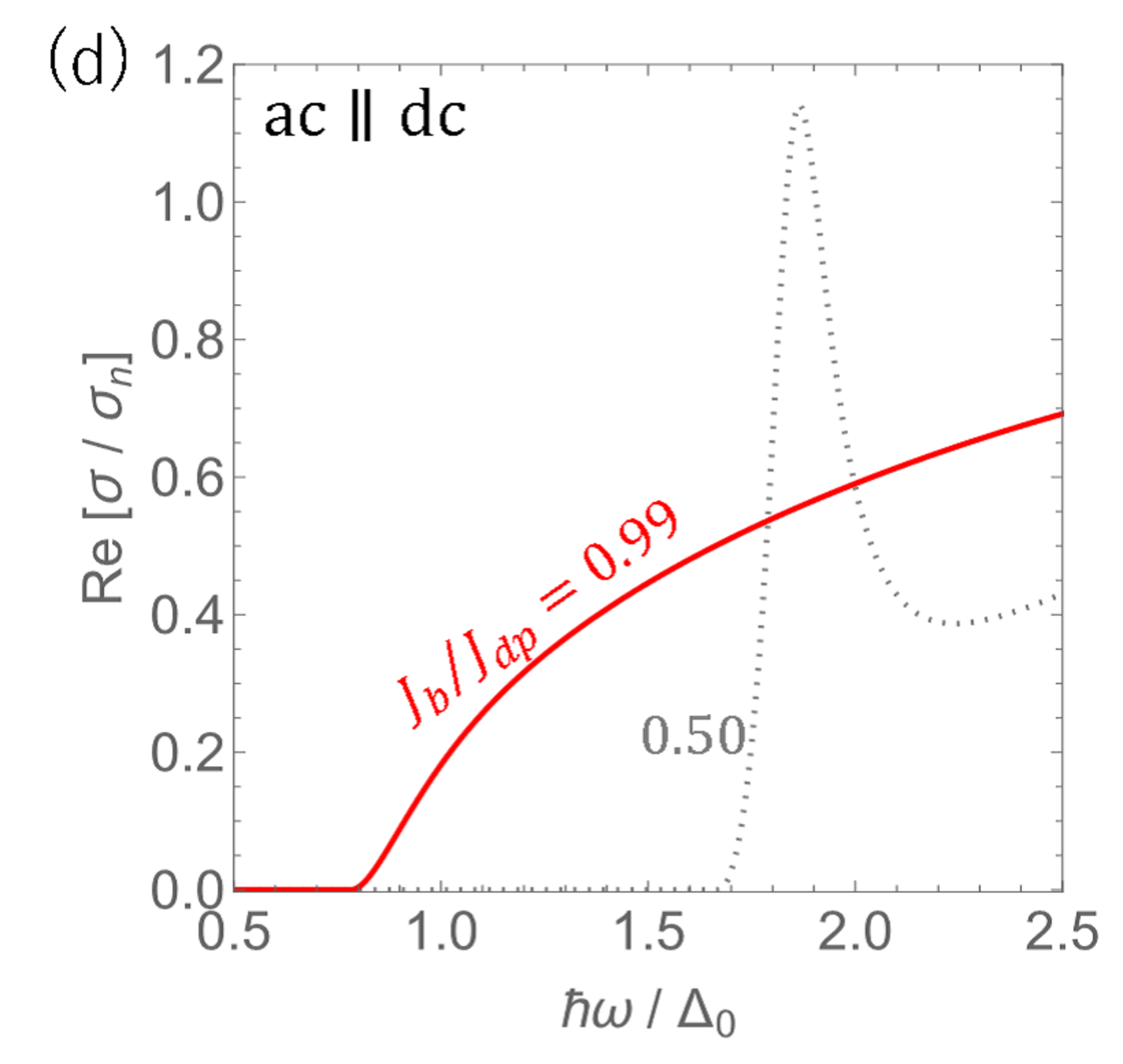}
   \includegraphics[width=0.24\linewidth]{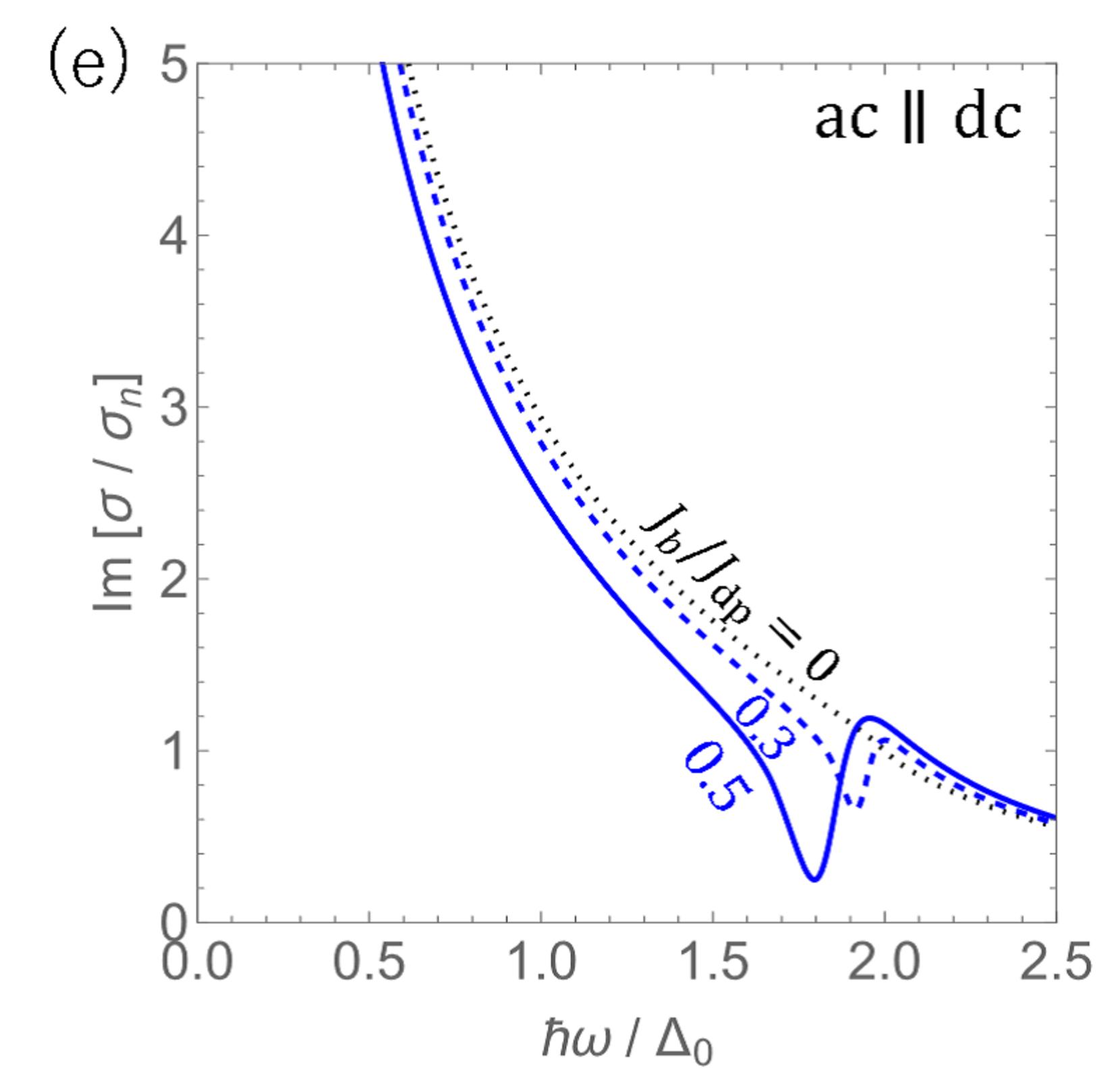}
   \includegraphics[width=0.24\linewidth]{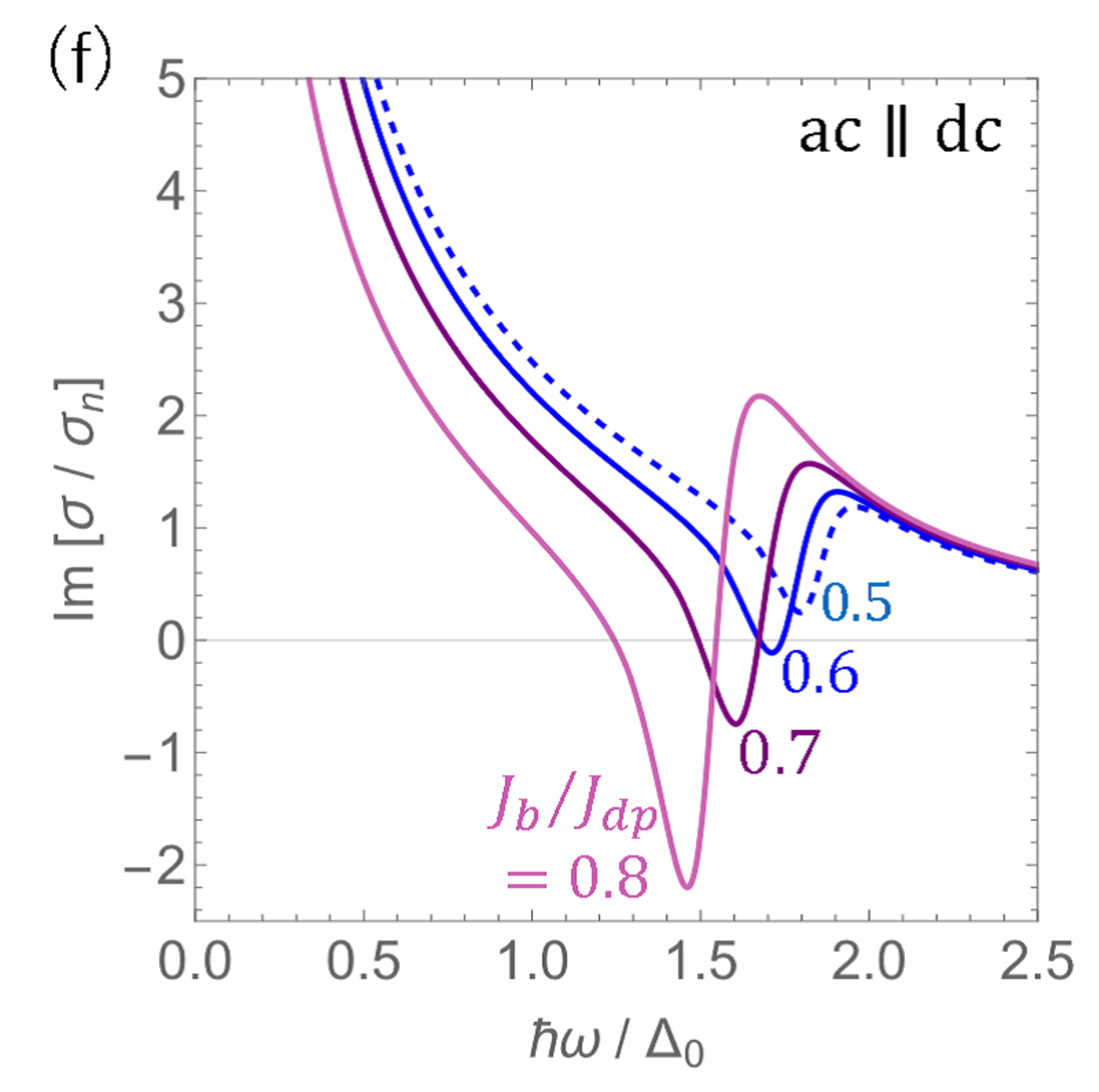}
   \includegraphics[width=0.24\linewidth]{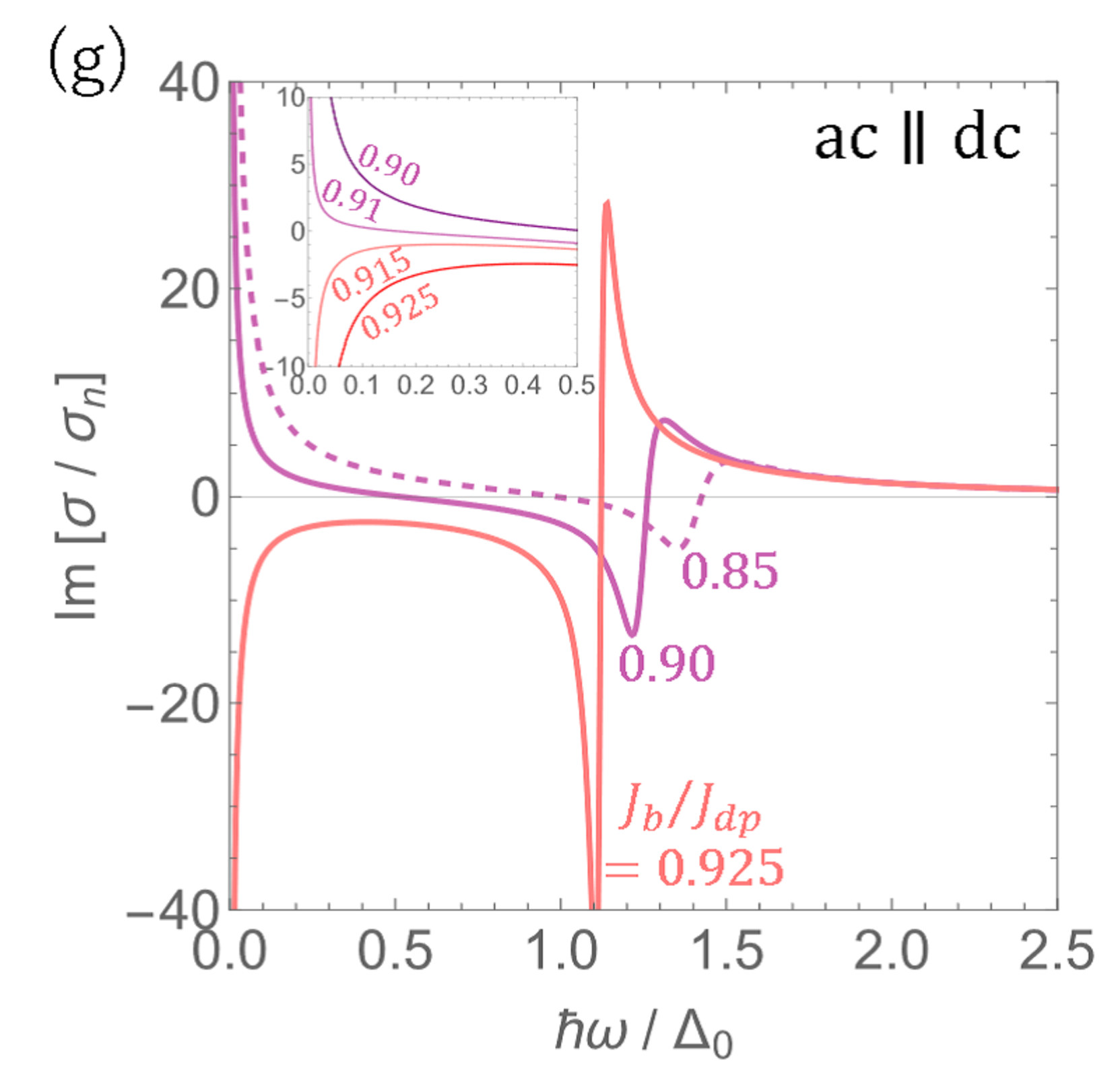}
   \includegraphics[width=0.24\linewidth]{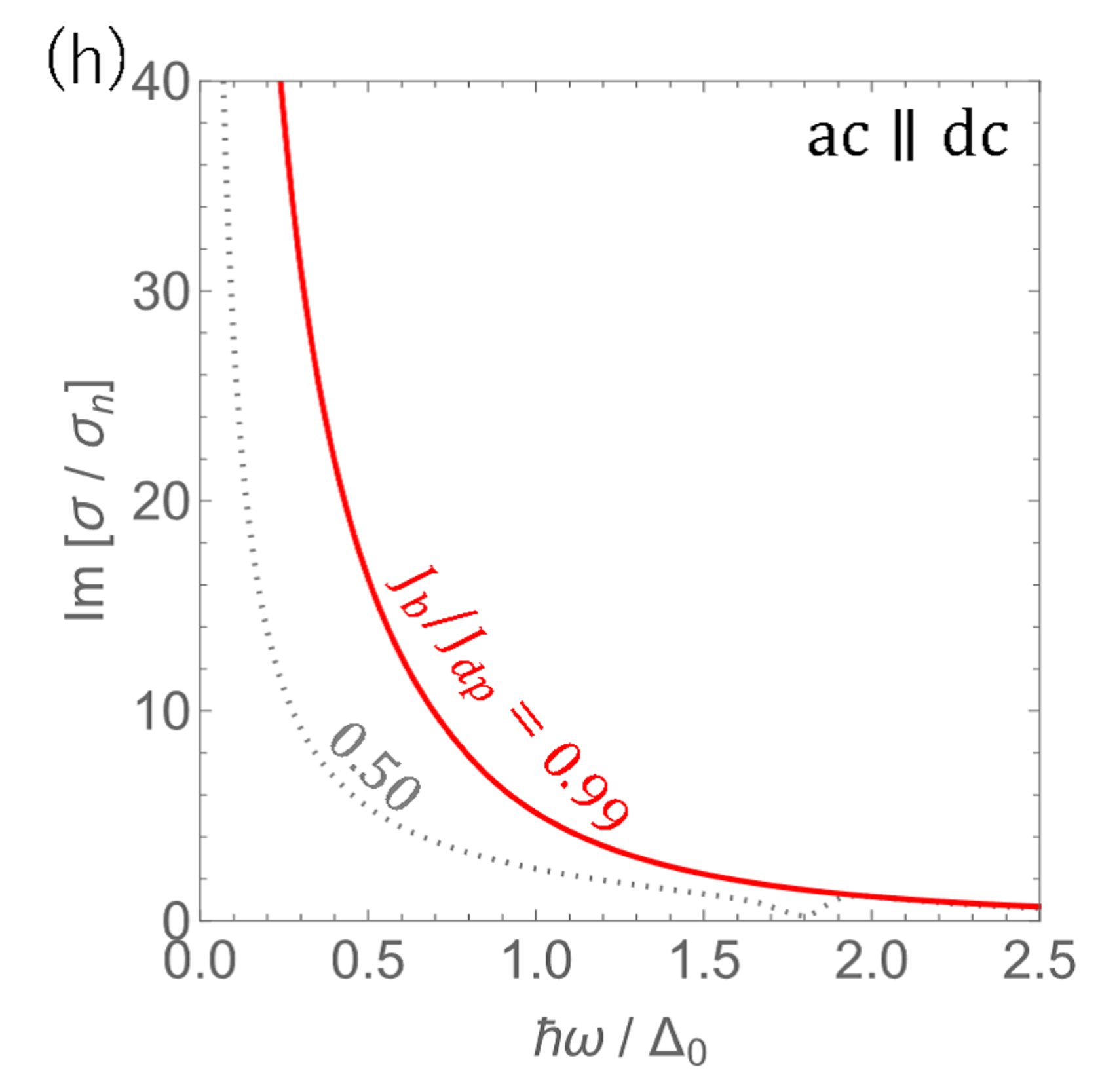}
   \end{center}\vspace{0 cm}
   \caption{
Nonperturbative Higgs mode manifestation in the complex conductivity for the \( {\rm ac} \parallel {\rm dc} \) configuration under various dc bias strengths.  
(a-d) Real and (e-h) imaginary parts of the complex conductivity, calculated at \( T = 0 \).  
As the bias dc increases, the Higgs resonance becomes more pronounced, leading to nontrivial frequency and bias dependencies, including the onset of instability at sufficiently strong bias currents.
   }\label{fig5}
\end{figure*}

Figure~\ref{fig5} presents the real (a-d) and imaginary (e-h) parts of the complex conductivity for the \( {\rm ac} \parallel {\rm dc} \) configuration, calculated at \( T=0 \).  
Figures~\ref{fig5}(a) and \ref{fig5}(e) correspond to moderately strong dc biases, where the Higgs resonance appears as a peak in \( {\rm Re} \, \sigma \) and a dip in \( {\rm Im} \, \sigma \), consistent with the perturbative results shown in Fig.~\ref{fig3}.  
These features become more pronounced as the bias strength increases.  
However, for even stronger dc biases, the Higgs mode contribution leads to highly nontrivial dependencies on both frequency and bias strength, as discussed below.

Figures~\ref{fig5}(b) and \ref{fig5}(f) illustrate cases with stronger dc biases, ranging from \( J_b / J_{\rm dp} = 0.5 \) to 0.8, where the Higgs resonance drives \( {\rm Im} \, \sigma \) strongly negative.  
For \( J_b / J_{\rm dp} \gtrsim 0.6 \), this effect results in two characteristic frequencies:  
\begin{eqnarray}
\omega_1(J_b), \omega_2(J_b) \in \{ \omega \mid {\rm Im} \, \sigma (J_b, \omega) = 0 \}. 
\end{eqnarray}
Within the frequency range \( \omega_1(J_b) < \omega < \omega_2(J_b) \), \( {\rm Im} \, \sigma \) (or equivalently, the superfluid density) becomes negative.  
It is well known that a negative \( {\rm Im} \, \sigma \) (or superfluid density) signifies instability, as the kinetic energy of the superflow decreases with increasing superfluid momentum.  
In this system, the homogeneous solutions of the Keldysh-Usadel equations become unstable under an ac perturbation within this instability frequency range, potentially leading to phase slips or vortex nucleation.

Figures~\ref{fig5}(c) and \ref{fig5}(g) illustrate cases with even stronger dc biases, ranging from \( J_b / J_{\rm dp} = 0.85 \) to 0.925.  
For \( J_b / J_{\rm dp} = 0.85 \) and 0.90, the frequency dependence remains similar to that observed for \( J_b / J_{\rm dp} \lesssim 0.8 \).  
However, for \( J_b / J_{\rm dp} = 0.925 \), the lower bound of the instability frequency window, \( \omega_1 \), disappears, and \( {\rm Im} \, \sigma (\omega) \) remains negative across the entire low-frequency range below \( \omega_2 \).  
This disappearance of \( \omega_1 \) indicates that the homogeneous current-carrying state becomes unstable against arbitrarily low-frequency ac perturbations as the bias current approaches \( J_b / J_{\rm dp} \simeq 0.9 \).

As the bias strength increases further, the upper bound of the instability frequency window, \( \omega_2 \), shifts to lower frequencies and eventually disappears, leading to the complete closure of the instability window.  
Consequently, the homogeneous current-carrying state regains stability against ac perturbations.  
Figures~\ref{fig5}(d) and \ref{fig5}(h) illustrate this stabilization for \( J_b / J_{\rm dp} = 0.99 \), with the results for \( J_b / J_{\rm dp} = 0.50 \) included for comparison.

Interestingly, despite the complex frequency and bias dependence, the onset of \( {\rm Re} \, \sigma \) in Figures~\ref{fig5}(a)-(d) is still given by \( \hbar \omega = 2\epsilon_g \), where \( \epsilon_g(J_b) \) is shown in Fig.~\ref{fig2}(b).

\begin{figure}[tb]
   \begin{center}
   \includegraphics[width=0.8\linewidth]{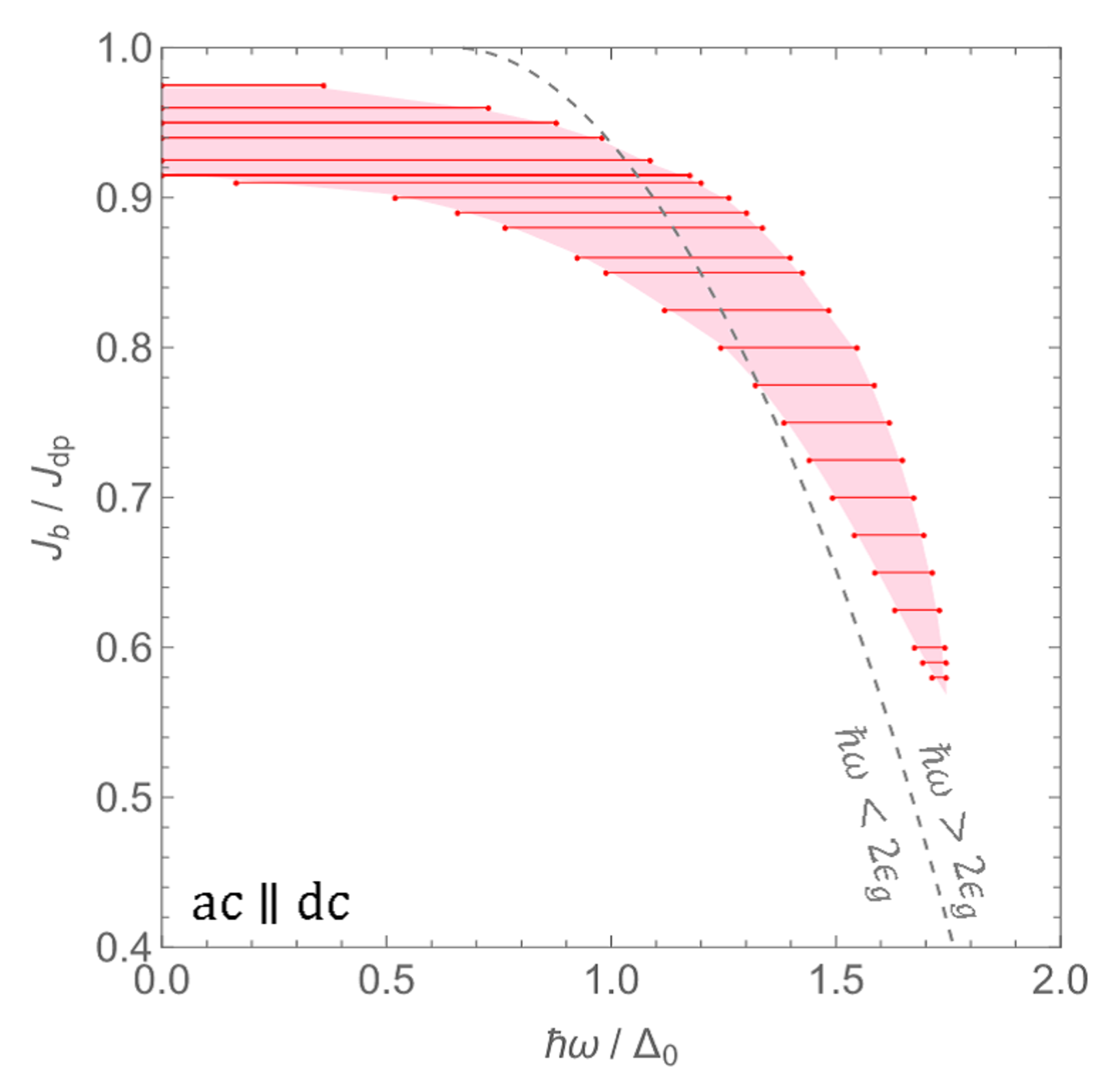}
   \end{center}\vspace{0 cm}
   \caption{
Higgs-induced instability domain in the \( \omega \)-\( J_b \) plane.  
The colored region represents the parameter space where \( {\rm Im} \, \sigma(J_b, \omega) < 0 \), indicating that the homogeneous current-carrying state is unstable against ac perturbations.  
Within this instability region, the system may transition into an inhomogeneous state, leading to phase slips or vortex nucleation.  
For comparison, the dashed curve represents the spectral gap \( 2\epsilon_g(J_b) \).  
   }\label{fig6}
\end{figure}

Figure~\ref{fig6} summarizes the Higgs-induced instability domain in the \( \omega \)-\( J_b \) plane, based on the calculations presented above.  
This instability, driven by the Higgs mode, occurs exclusively in the \( {\rm ac} \parallel {\rm dc} \) configuration and is entirely absent in the \( {\rm ac} \perp {\rm dc} \) configuration.  
Further discussion on this topic is provided in Section~\ref{Discussion_B}.

In the next section, we focus on bias-dependent kinetic inductance, a key quantity in superconducting device applications.  
While the primary emphasis is on the kinetic inductance, the Higgs-induced instability plays a crucial role in its behavior.  
Thus, we revisit this instability within the kinetic inductance framework, offering a more intuitive perspective that enhances comprehension.

%%%%%%%%%%%%%%%%%%%
%%%%%%%%%%%%%%%%%%%
%%%%%%%%%%%%%%%%%%%
\section{Bias-dependent kinetic inductance}\label{section_Lk}
%%%%%%%%%%%%%%%%%%%
%%%%%%%%%%%%%%%%%%%
%%%%%%%%%%%%%%%%%%%

%%%%%%%%%%%%%%%%%%%
\subsection{Kinetic inductivity}
%%%%%%%%%%%%%%%%%%%

The kinetic inductivity is defined by $L_k \dot{J}_s = E$, or equivalently, in the frequency domain, $L_k \{ -i\omega J_s(\omega) \} = E(\omega)$.  
Here, the reactive component of the current density is given by $J_s = i \sigma_2 E$ and $\sigma_2:={\rm Im} \sigma$. 
Thus, the kinetic inductivity can be expressed as:
\begin{eqnarray}
L_k 
&=& \frac{1}{\omega \sigma_2} \label{Lk} \\
&=& 
\begin{cases}
\big\{ \omega (\sigma_2^{(0)} +\sigma_2^{(1)} + \sigma_2^{(2)} ) \big\}^{-1} & ({\rm ac} \parallel {\rm dc}) \\
\big( \omega \sigma_2^{(0)} \big)^{-1} & ({\rm ac} \perp {\rm dc}) 
\end{cases} \label{Lk2}
\end{eqnarray}
Here, $\sigma_2^{(i)}:={\rm Im} \sigma^{(i)}$ ($i=0, 1, 2$).
For example, in the zero-bias ($s = 0$), low-frequency ($\hbar\omega \ll \Delta_0$), and low-temperature ($kT \ll \Delta_0$) limit, translating the integral over $\epsilon$ into a Matsubara summation yields $\sigma_{20}:= \sigma_2(0, 0, 0) = \sigma_2^{(0)}(0, 0, 0) = 1 / \mu_0 \omega \lambda_0^2= \pi \Delta_0 \sigma_n/\hbar \omega$, reproducing the well-known result $L_{k0} = 1/\omega \sigma_{20}= \mu_0 \lambda_0^2$.

%%%%%%%%%%%%%%%%%%%
\subsection{Behavior in the Small Bias regime}\label{weak_dc_Lk}
%%%%%%%%%%%%%%%%%%%

While our formulation applies to arbitrary bias dc strengths, particular insight can be gained by examining the small bias regime (\( J_b \ll J_{\rm dp} \)).  
In this regime, the kinetic inductance follows Eq.~(\ref{Lk_expansion}), where the coefficient \( C \) plays a crucial role in distinguishing between the oscillating and frozen superfluid density regimes.  
As we will demonstrate later in this subsection, these regimes correspond to the \( {\rm ac} \parallel {\rm dc} \) and \( {\rm ac} \perp {\rm dc} \) configurations, respectively.

Evaluating $C$ is straightforward. Using Eq.~(\ref{Lk}), we can rewrite 
\begin{eqnarray}
\frac{L_k(J_b)}{L_k(0)} = 1 + C \biggl( \frac{J_b}{J_{\rm dp}} \biggr)^2,
\end{eqnarray}
as
\begin{eqnarray}
\frac{\sigma_2(J_b)}{\sigma_{2}(0)} = 1 - C \biggl( \frac{J_b}{J_{\rm dp}} \biggr)^2,
\end{eqnarray}
which is valid in the small bias regime. By calculating $\sigma_2$ and incorporating the effects of the bias dc up to order $s / \Delta_0 \sim (J_b / J_{\rm dp})^2$, we can determine the coefficient $C$.

We begin by focusing on the low-frequency ($\hbar\omega \ll \Delta$) and low-temperature ($T \ll T_c$) regime.  
This focus is motivated by two key reasons.  
First, in this regime, the coefficient $C$ can be calculated analytically.  
Second, many superconducting materials used in practical devices operate at technologically relevant frequencies ($\sim \mathrm{GHz}$), which generally satisfy $\hbar \omega \ll \Delta_0$, and their operating temperatures are typically in the low-temperature limit ($T \ll T_c$).  
Thus, the low-frequency and low-temperature regime is of both theoretical and practical importance.

The calculation of \( \sigma_2^{(0)} := \mathrm{Im} \, \sigma^{(0)} \) proceeds as follows.  
In the low-frequency limit, \( \sigma_2^{(0)} \), given by Eq.~(\ref{sigma0}), reduces to the Matsubara sum \( (4\pi kT \sigma_n / \hbar \omega) \sum_{\omega_m > 0} F_m^2 \).  
Using Eq.~(\ref{rho_s}), this evaluates to \( 1 / \mu_0 \omega \lambda^2(s, T) \) for any bias and temperature.  
For $T \to 0$, the Matsubara summation $2\pi k T \sum_{\omega_m > 0} (\dots)$ can be replaced by the integral $\Delta_0 \int_0^{\infty} (\dots) dw$, where $w = \hbar \omega_m / \Delta_0$. Performing this substitution yields:
\begin{eqnarray}
\frac{\sigma_2^{(0)}(s, \omega, T)}{\sigma_{20}}\Bigr|_{\omega \ll \frac{\Delta_0}{\hbar}, \, T \ll T_c} 
= 1 - \Bigl( \frac{\pi}{4} + \frac{4}{3\pi} \Bigr) \frac{s}{\Delta_0} 
+ \mathcal{O}(s^2). \nonumber \\
\end{eqnarray}

The calculations of \( \sigma_2^{(1,2)} := {\rm Im} \, \sigma^{(1,2)} \) are more straightforward.  
Since this subsection considers the effects of the bias dc up to the order of \( \mathcal{O}(s) \), we can substitute the zero-current Green's functions in place of \( G_b \) and \( F_b \) in Eqs.~(\ref{sigma1}) and (\ref{sigma2}), leading to 
\begin{eqnarray}
&&\frac{\sigma_2^{(1)}(s,\omega,T)}{\sigma_{20}}\Bigr|_{\omega\ll\frac{\Delta_0}{\hbar}, \,T \ll T_c} = -\frac{8}{3\pi}\frac{s}{\Delta_0} + \mathcal{O}(s^2) , \\
&&\frac{\sigma_2^{(2)}(s,\omega,T)}{\sigma_{20}}\Bigr|_{\omega\ll\frac{\Delta_0}{\hbar}, \,T \ll T_c} = \Psi \frac{s}{\Delta_0} + \mathcal{O}(s^2) , 
\end{eqnarray}
Here, $\Psi$ for $\hbar \omega \ll \Delta_0$ can also be evaluated analytically in a similar manner.  
In the numerator, the integral simplifies to $-2\pi$ as $T \to 0$.  
For the denominator, the integral simplifies to \( 4(\sinh^{-1} \tilde{\omega}_c - 1) \), where \( \tilde{\omega}_c = \hbar \omega_c / \Delta_0 \) is the dimensionless cutoff energy of the Matsubara sum, with \( \omega_c \) typically set near the Debye frequency.  
The BCS coupling constant ${\mathscr G}$ can be eliminated by utilizing the BCS relation $1 = {\mathscr G} \sinh^{-1} \tilde{\omega}_c$, yielding 
\begin{eqnarray}
\Psi(s,\omega,T)\bigr|_{\omega\ll\frac{\Delta_0}{\hbar}, \,T \ll T_c} = -\frac{\pi}{2} .
\end{eqnarray}

The next step is to express $s / \Delta_0$ in terms of $J_b / J_{\rm dp}$.  
In general, as illustrated in Fig.~\ref{fig2}(a), $J_b / J_{\rm dp}$ exhibits a nonlinear dependence on $s / \Delta_0 = (q_b / q_{\xi})^2$.  
However, in the small bias regime ($J_b / J_{\rm dp} \ll 1$), this relationship simplifies to a linear form: $J_b / J_{\rm dp} = k q_b / q_{\xi}$, which leads to $s/\Delta_0=k^{-2} (J_b/J_{\rm dp})^2$. 
While the slope $k(T)$ generally needs to be determined numerically for a given temperature $T$, its analytical expression at $T \to 0$ is derived in the Appendix as:
\begin{eqnarray}
k(0) = \frac{1}{\sqrt{s_d / \Delta_0} \bigl(\Delta_d / \Delta_0 - 4 s_d / 3\pi \Delta_0\bigr)} \simeq 2.98.
\end{eqnarray}
Here, $\eta_d = 2\pi + 3\pi/8 - \sqrt{(2 / \pi + 3\pi/8)^2 - 1} = 0.300$, $\Delta_d / \Delta_0 = e^{-\pi \eta_d / 4} = 0.790$, and $s_d / \Delta_0 = \eta_d (\Delta_d / \Delta_0) = 0.237$.

Finally, we arrive at the analytical expression for $C$ in the low-frequency and low-temperature regime.  
The result is:
\begin{eqnarray}
C &=&
\begin{cases}
C^{(0)} + C^{(1)} + C^{(2)}  & ({\rm ac} \parallel {\rm dc})  \\
C^{(0)}  & ({\rm ac} \perp {\rm dc}) 
\end{cases} \nonumber \\
&\simeq & 
\begin{cases}
0.409 & ({\rm ac} \parallel {\rm dc})  \\
0.136 & ({\rm ac} \perp {\rm dc}) 
\end{cases} .
\end{eqnarray}
Here, the contributions are:
\begin{eqnarray}
&& C^{(0)} = \Bigl( \frac{\pi}{4} + \frac{4}{3\pi} \Bigr) \frac{s_d}{\Delta_0} \biggl( \frac{\Delta_d}{\Delta_0} - \frac{4 s_d}{3\pi \Delta_0} \biggr)^2 \simeq 0.136, \\ 
&& C^{(1)} = \frac{8}{3\pi} \frac{s_d}{\Delta_0} \biggl( \frac{\Delta_d}{\Delta_0} - \frac{4 s_d}{3\pi \Delta_0} \biggr)^2 \simeq 0.0956, \\ 
&& C^{(2)} = \frac{\pi}{2} \frac{s_d}{\Delta_0} \biggl( \frac{\Delta_d}{\Delta_0} - \frac{4 s_d}{3\pi \Delta_0} \biggr)^2 \simeq 0.177.  
\end{eqnarray}
It follows that $C^{(1)} + C^{(2)} = 2C^{(0)}$. 
The above results indicate that more than 40\% of the bias dependence in the \( {\rm ac} \parallel {\rm dc} \) configuration originates from the contribution of the Higgs mode.

The resulting values of \( C \) for the \({\rm ac} \parallel {\rm dc}\) and \({\rm ac} \perp {\rm dc}\) configurations precisely match the previous calculations based on the thermodynamic Usadel equation combined with the oscillating and frozen superfluid density assumptions, respectively~\cite{2020_Kubo, 2020_Kubo_erratum}. 
Notably, while a prior study by the author~\cite{2024_Kubo} identified these coincidences through numerical calculations based on the Keldysh-Eilenberger equation, we now confirm them analytically.

It should be noted that the above coefficient \( C \) is evaluated in the dirty limit at \( T=0 \).  
For different mean free paths and temperatures, \( C \) varies, and its numerical evaluation can be found in Ref.~\cite{2024_Kubo}.

%%%%%%%%%%%%%%%%%%%
\subsection{Kinetic Inductance Under Arbitrary Bias Strength and Higgs-Induced Instability} \label{strong_dc_Lk}
%%%%%%%%%%%%%%%%%%%

The calculation of bias-dependent \( L_k \) for an arbitrary bias dc strength [see Eq.~(\ref{Lk})] follows the same procedure as the evaluation of \( {\rm Im} \, \sigma \) presented in Figures~\ref{fig4}-\ref{fig6}.

\begin{figure}[tb]
   \begin{center}
   \includegraphics[height=0.48\linewidth]{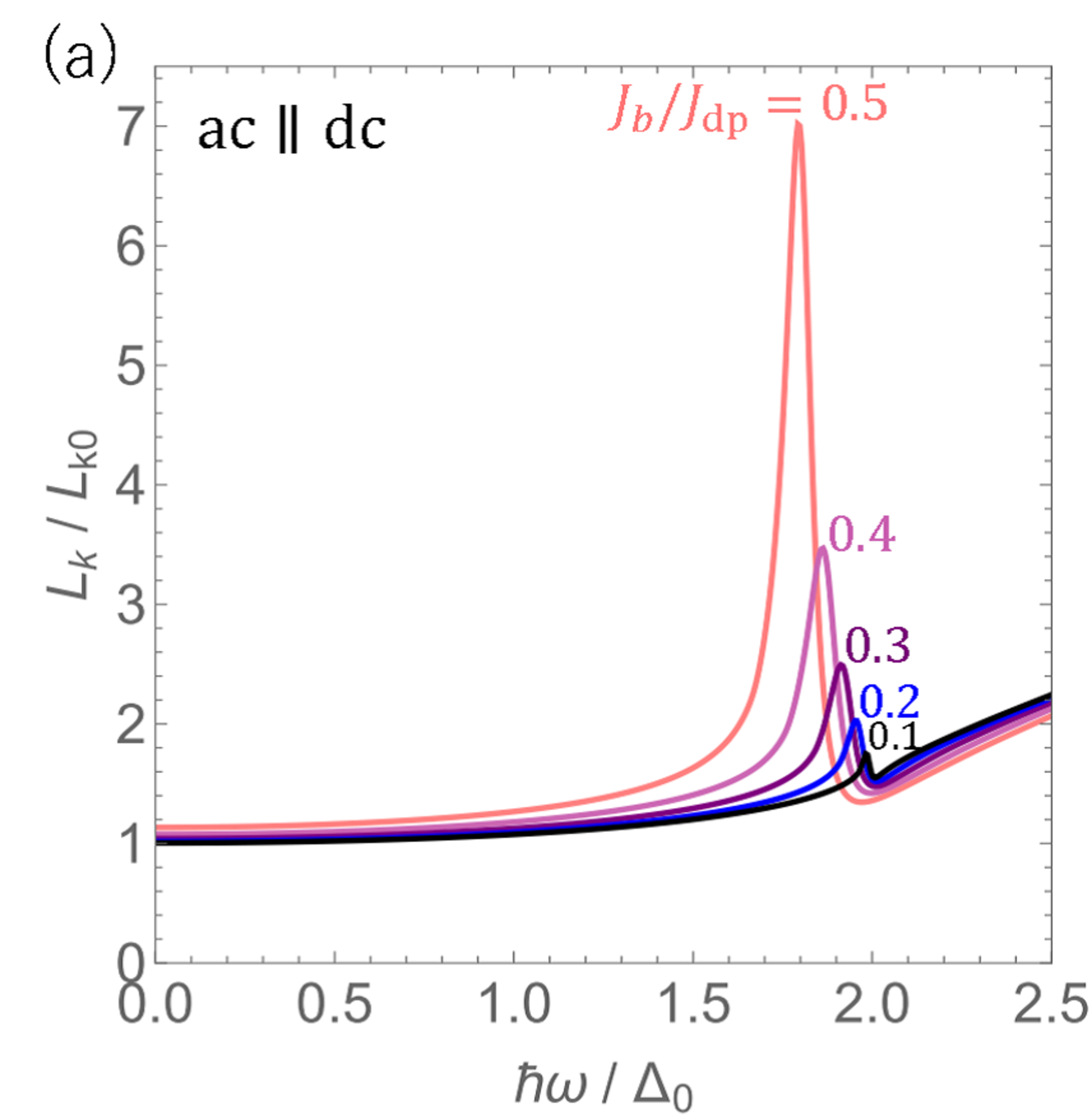}
   \includegraphics[height=0.48\linewidth]{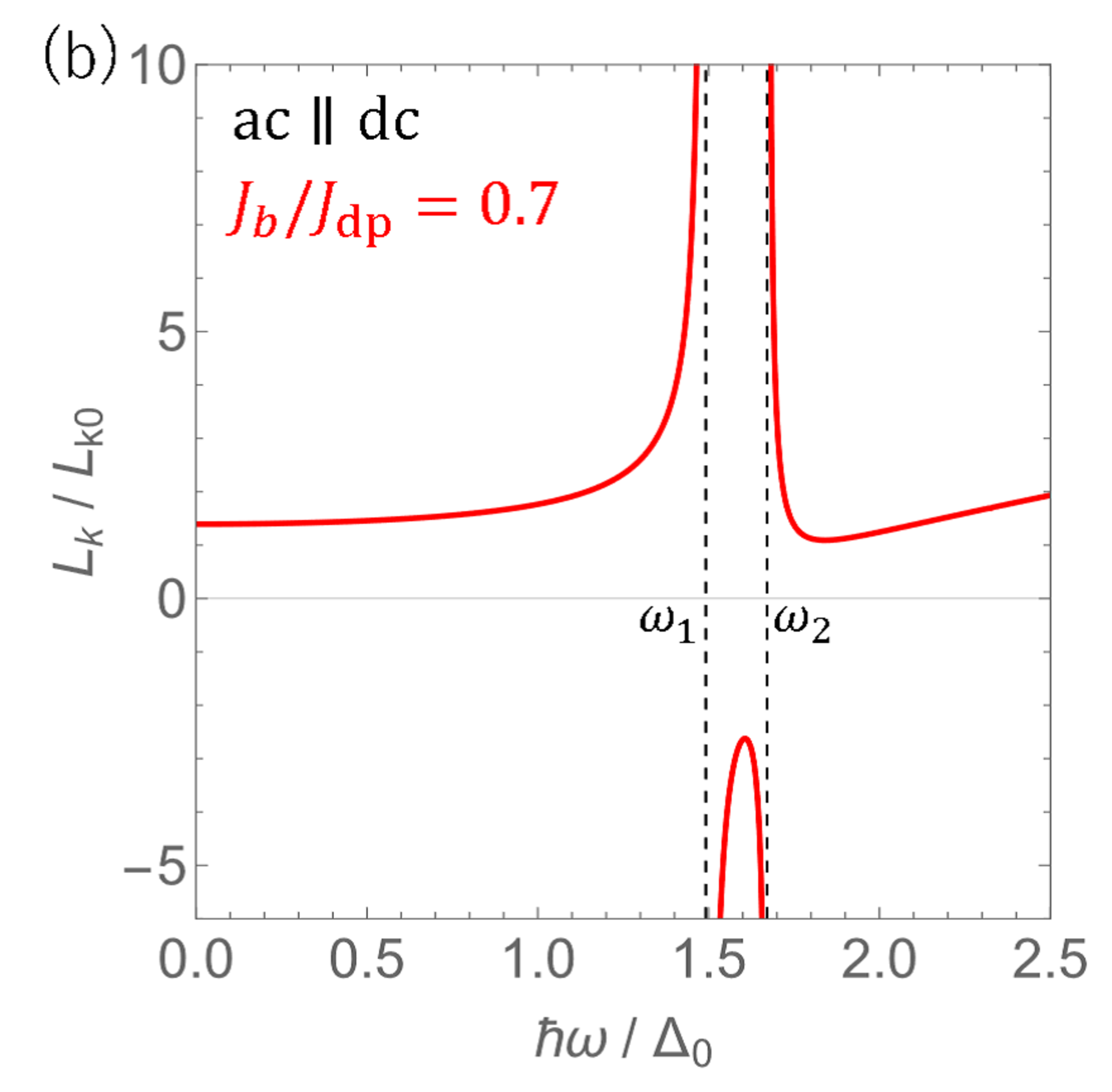}
   \includegraphics[height=0.47\linewidth]{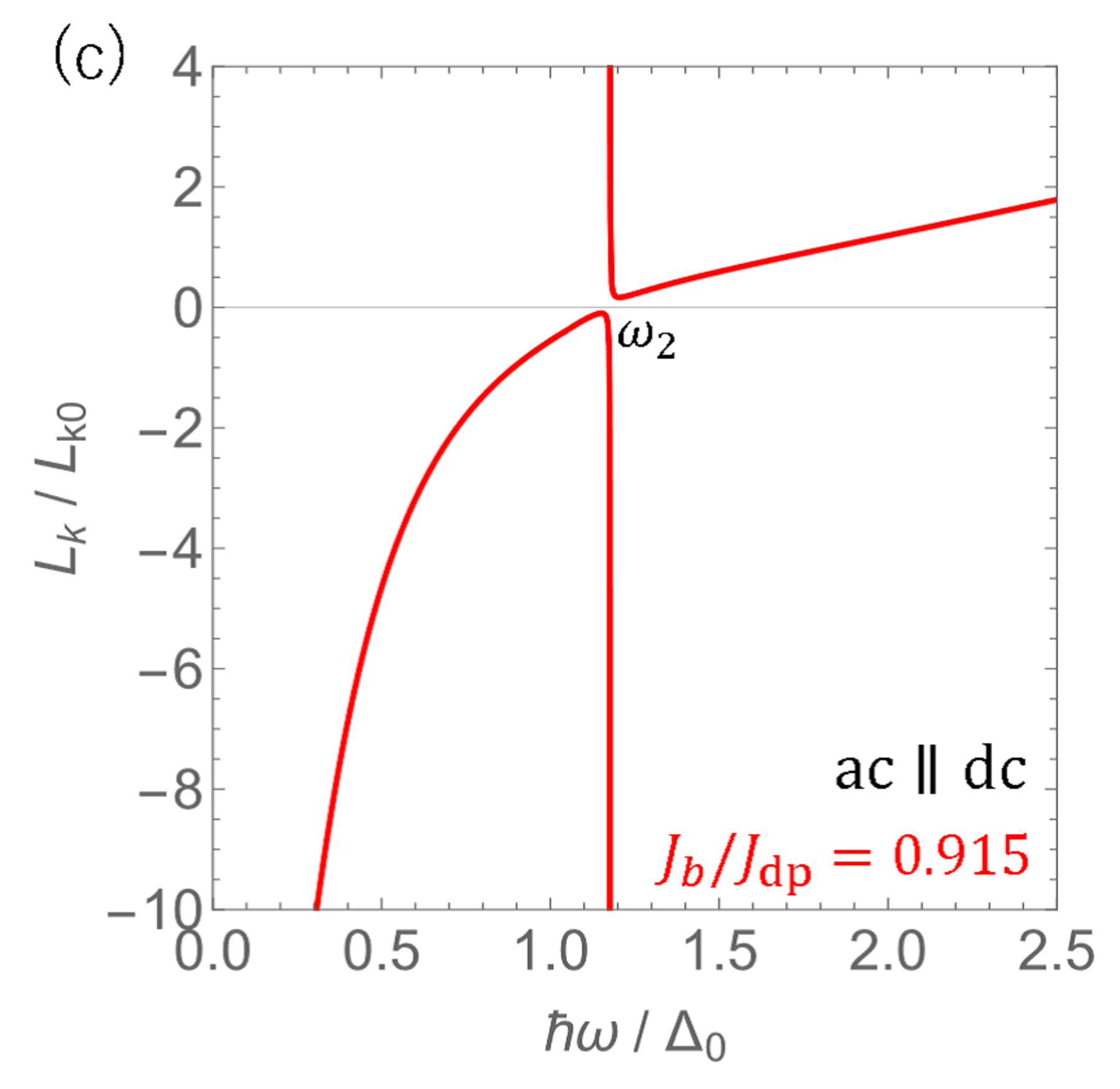}
   \includegraphics[height=0.47\linewidth]{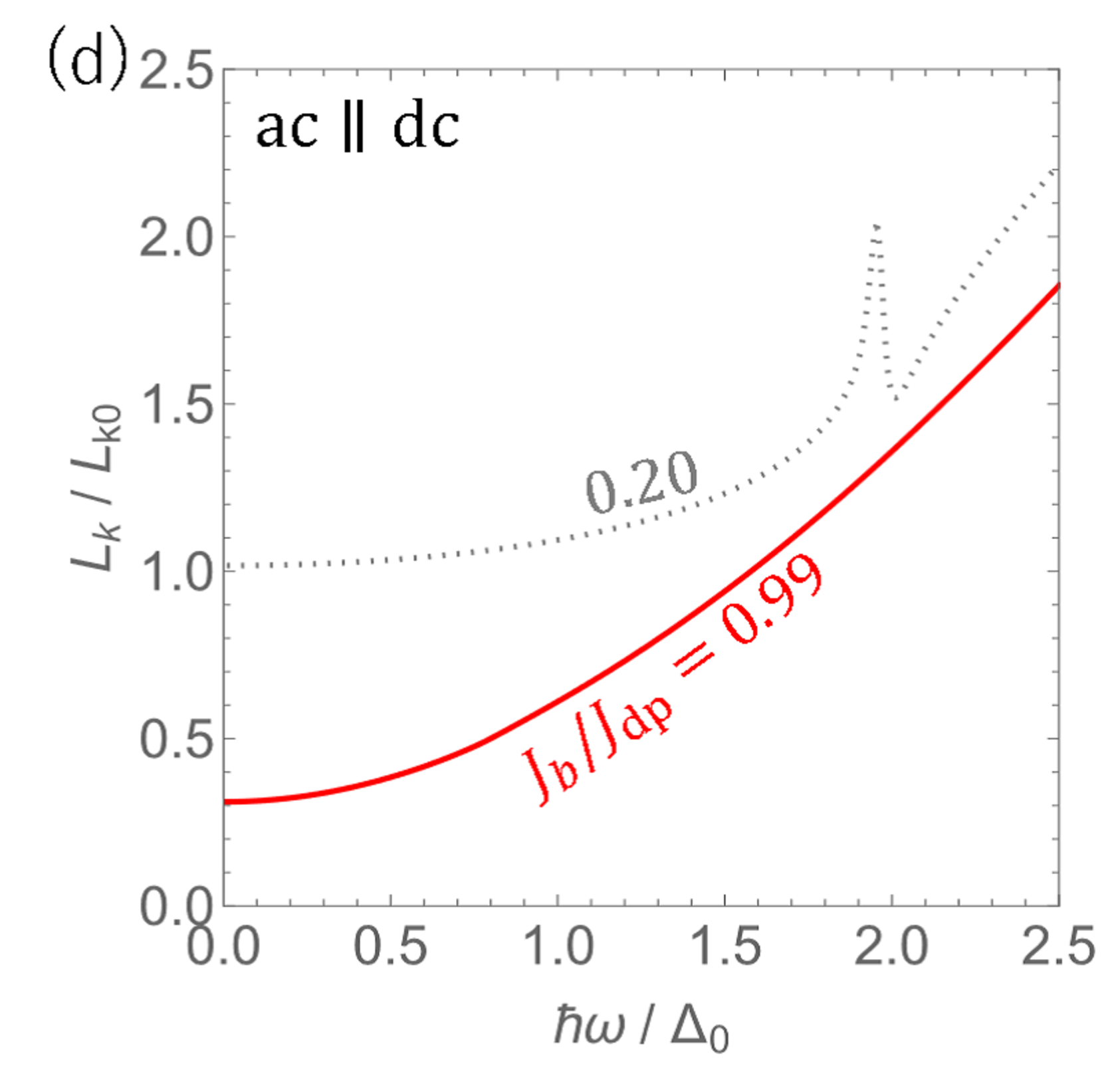}
   \end{center}\vspace{0 cm}
   \caption{
Frequency dependence of the kinetic inductance \( L_k \) at \( T=0 \) for the \( {\rm ac} \parallel {\rm dc} \) configuration, calculated at various dc biases:  
(a) \( J_b/J_{\rm dp} = 0.1 \) to \( 0.5 \), (b) \( J_b/J_{\rm dp} = 0.7 \), (c) \( J_b/J_{\rm dp} = 0.915 \), and (d) \( J_b/J_{\rm dp} = 0.99 \).  
The occurrence of negative \( L_k \) indicates the instability of the homogeneous current-carrying state, leading to a transition into an inhomogeneous state, such as phase slips or vortex nucleation.  
It is important to note that in real experiments, rather than directly observing negative values, \( L_k \) corresponding to these inhomogeneous states is expected to be measured, along with a finite voltage generated by moving vortices.
   }\label{fig7}
\end{figure}

Figure~\ref{fig7}(a) illustrates the frequency dependence of \( L_k (\omega) \) for small to moderately strong bias dc values in the \( {\rm ac} \parallel {\rm dc} \) configuration.  
In this regime, \( L_k (\omega) \) exhibits weak frequency dependence, except around \( \hbar \omega \simeq 2\Delta \), where a pronounced peak emerges due to the Higgs mode contribution.  
This peak directly corresponds to the dip observed in \( {\rm Im} \, \sigma \) [see Figure~\ref{fig6}(e)].  
It is worth noting that, in the \( {\rm ac} \perp {\rm dc} \) configuration, the Higgs-induced peak vanishes, resulting in \( L_k (\omega) \) exhibiting a monotonically increasing trend with weak frequency dependence [see also Figure~\ref{fig4}].

As the bias dc strength increases further (\( J_b/J_{\rm dp} \gtrsim 0.6 \)), as shown in Figure~\ref{fig7}(b), the peak in \( L_k \) diverges at the characteristic frequencies \( \omega_1 \) and \( \omega_2 \), where \( {\rm Im} \, \sigma = 0 \).  
The frequency range \( \omega_1 < \omega < \omega_2 \), where \( L_k < 0 \), marks the onset of Higgs-induced instability in the homogeneous current-carrying state under ac perturbations, potentially leading to phase slips or vortex nucleation (see also the discussion in Section~\ref{strong_dc_sigma}).

With further increase in the bias strength [see Figure~\ref{fig7}(c)], the instability window shifts to lower frequencies, eventually reaching \( \omega_1 \to 0 \), where the instability domain spans the range \( 0< \omega < \omega_2 \).  

For even stronger dc bias [see Figure~\ref{fig7}(d)], \( \omega_2 \) also shifts toward zero. 
At this stage, the homogeneous current-carrying state regains stability across the entire ac spectrum.  
Interestingly, the kinetic inductance in this regime is found to be smaller than that in the zero-current state.

\begin{figure}[tb]
   \begin{center}
   \includegraphics[height=0.45\linewidth]{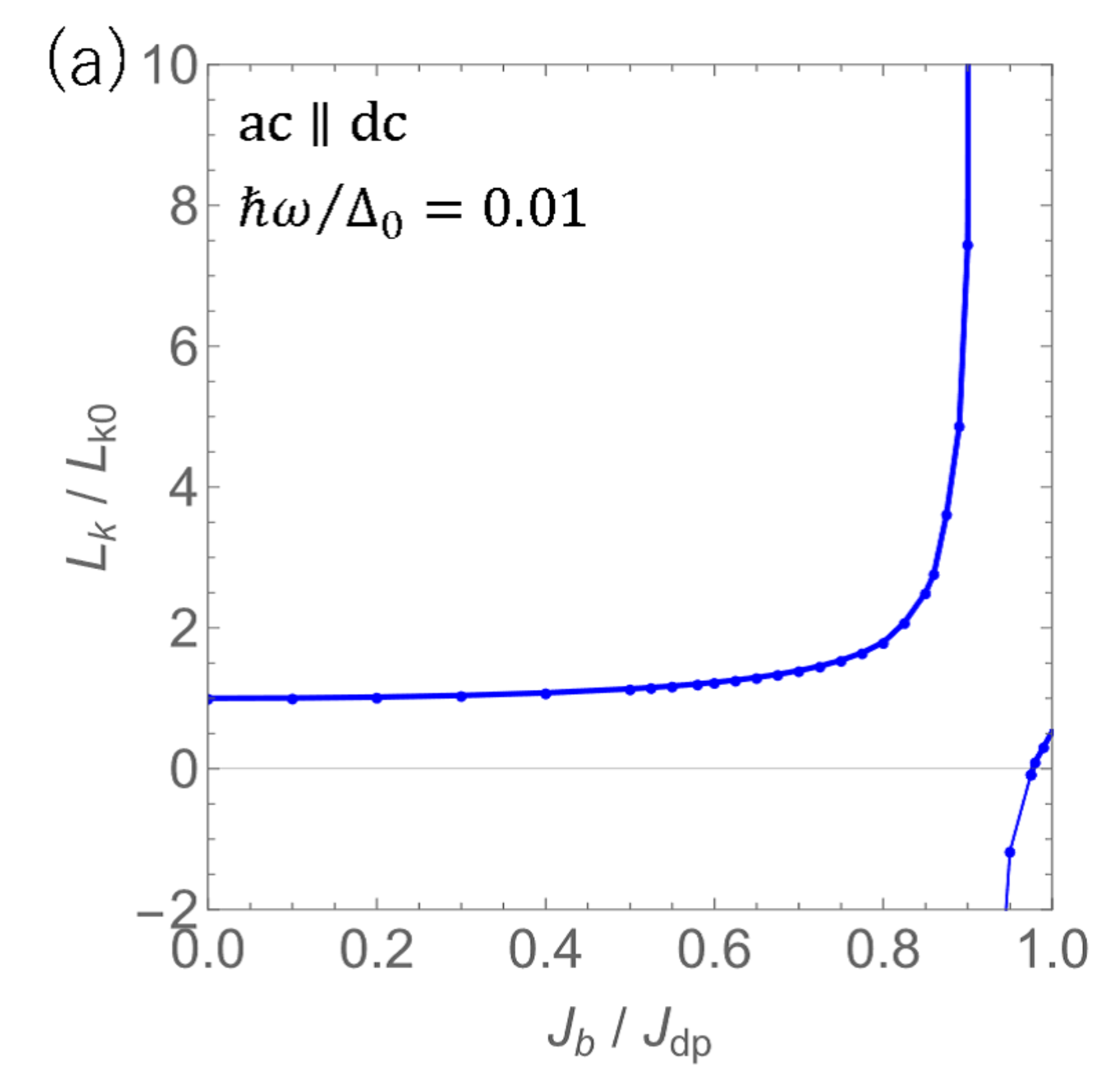}
   \includegraphics[height=0.45\linewidth]{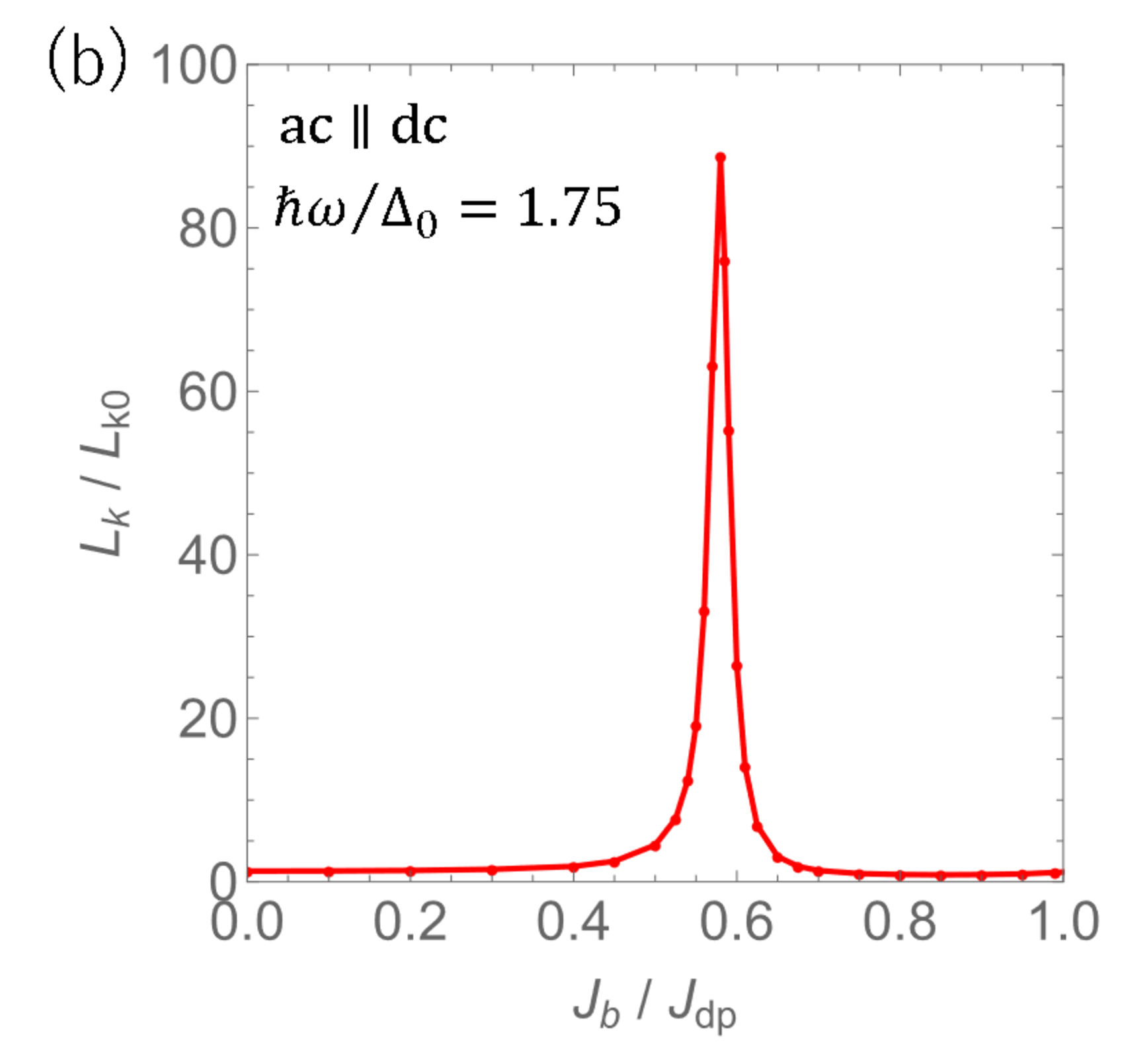}
   \end{center}\vspace{0 cm}
   \caption{
Bias dc dependence of the kinetic inductance \( L_k \) for the \( {\rm ac} \parallel {\rm dc} \) configuration, calculated at \( T=0 \).  
The frequency of the ac perturbation is  
(a) \( \hbar\omega /\Delta_0 =0.01 \) and (b) \( \hbar\omega /\Delta_0 =1.75 \).  
Negative \( L_k \) signals the instability of the homogeneous current-carrying state, leading to a transition into an inhomogeneous state with phase slips or vortex nucleation.  
In real experiments, rather than observing negative values, an effective \( L_k \) associated with these inhomogeneous states is expected to be measured.  
   }\label{fig8}
\end{figure}

For completeness, we summarize the above findings in terms of bias dependence.  
Figure~\ref{fig8}(a) presents \( L_k(J_b) \) for a low-frequency ac perturbation (\(\hbar\omega /\Delta_0 =0.01\)).  
As the dc bias increases, \( L_k \) initially rises and diverges to \( +\infty \), before abruptly dropping to negative values around \( J_b/J_{\rm dp} \simeq 0.9 \), signaling the Higgs-induced instability of the homogeneous current-carrying state (see the left edge of Fig.~\ref{fig6}).  
For \( J_b/J_{\rm dp} \gtrsim 0.98 \), stability is restored, and \( L_k \) regains positive values.  
Interestingly, in this regime, \( L_k \) is smaller than that of the zero-current state.  
It is important to note that a negative \( L_k \) signifies instability rather than a physically observable quantity.  
In real experiments, instead of directly measuring negative \( L_k \), one would observe an effective \( L_k \) corresponding to the emergent inhomogeneous states, which include vortices and are accompanied by a finite voltage due to vortex motion.

Figure~\ref{fig8}(b) presents \( L_k(J_b) \) for an ac perturbation with frequency \( \hbar\omega /\Delta_0 =1.75 \).  
In this case, no negative \( L_k(J_b) \) appears, confirming the absence of instability (see also the corresponding frequency in Fig.~\ref{fig6}).  
Instead, a pronounced peak emerges, associated with a tiny \( {\rm Im} \, \sigma \) at the boundary of the instability domain.  
The maximum value of \( L_k \) occurs around half the depairing current and is nearly two orders of magnitude larger than its zero-current counterpart.

%%%%%%%%%%%%%%%%%%%
%%%%%%%%%%%%%%%%%%%
\section{Discussion}
%%%%%%%%%%%%%%%%%%%
%%%%%%%%%%%%%%%%%%%

In this paper, we have investigated a dirty-limit superconductor subjected to a perturbative ac field superposed on a dc bias of arbitrary strength.  
Our analysis has revealed that strong dc biases parallel to the ac field give rise to highly nontrivial effects due to the pronounced contribution of the Higgs mode in dirty superconductors, leading to striking phenomena that cannot be captured in the weak-bias regime.  
Below, we summarize our key findings and discuss their implications for applied superconductivity.

%%%%%%%%%%
\subsection{Complex conductivity formula}
%%%%%%%%%%

In Section~\ref{formulation}, we derived the complex conductivity formula for a disordered superconductor subjected to an ac perturbation superposed on a dc bias of arbitrary strength using the Keldysh-Usadel theory of nonequilibrium superconductivity.  
This formula, given by Eq.~(\ref{total_sigma}), represents one of the main results of this paper.  
In the weak dc bias regime, where the bias can be treated as a perturbation, our formula reproduces the results obtained by Moor et al.~\cite{Moor}.  
Furthermore, it serves as the dirty-limit counterpart of the more general expression derived from the Keldysh-Eilenberger theory in Refs.~\cite{Jujo, 2024_Kubo}.

It is important to note that previous studies investigating the ac response under a strong dc bias in the context of superconducting device applications (e.g., Refs.~\cite{Clem_Kogan, Gurevich_PRL, 2020_Kubo1, Zhao}) considered only the contribution from Eq.~(\ref{sigma0}) while neglecting the additional contributions from Eqs.~(\ref{sigma1}) and (\ref{sigma2}).  
However, their results remain valid within the specific context of the \( {\rm ac} \perp {\rm dc} \) configuration, where the Higgs mode is not excited.

For instance, Refs.~\cite{Gurevich_PRL, 2020_Kubo1} demonstrated that \( {\rm Re} \sigma \) exhibits a dc-bias-dependent reduction at low frequencies and at moderately low temperatures, suggesting the possibility of tuning dissipation via the bias dc.  
These findings hold for the \( {\rm ac} \perp {\rm dc} \) configuration, where the Higgs mode is not excited.  
In contrast, for the \( {\rm ac} \parallel {\rm dc} \) configuration, the additional contributions from Eqs.~(\ref{sigma1}) and (\ref{sigma2}) significantly modify \( {\rm Re} \sigma \), as shown in Ref.~\cite{2024_Kubo}.  
This results in a dc-bias-dependent suppression of \( {\rm Re} \sigma \) that extends to higher bias strengths and higher frequencies.

%%%%%%%%%%
\subsection{Higgs-induced instability of the homogeneous current-carrying state} \label{Discussion_B}
%%%%%%%%%%

Section~\ref{section_Higgs_and_sigma} demonstrated the Higgs-induced instability of the homogeneous current-carrying state, which arises in the \( {\rm ac} \parallel {\rm dc} \) configuration under a strong dc bias.  
As the bias increases beyond \( J_b / J_{\rm dp} \gtrsim 0.6 \), the Higgs mode contribution to \( {\rm Im} \, \sigma \) becomes strongly negative within a certain frequency range, leading to the emergence of a window where \( {\rm Im} \, \sigma < 0 \), indicating a negative superfluid density.  
This negative superfluid density implies that the kinetic energy decreases with increasing current, rendering the homogeneous current-carrying state unstable and likely to transition into an inhomogeneous state through vortex nucleation.  
This instability is mapped in the \( \omega \)-\( J_b \) plane in Fig.~\ref{fig6}.  

It should be noted that this instability map is specific to the dirty limit.  
As the mean free path increases, the Higgs-mode contribution to \( {\rm Im} \sigma \) is suppressed~\cite{2024_Kubo}, shifting the instability window to higher bias currents until it eventually disappears at a critical mean free path.  
Determining this critical mean free path is feasible using the theoretical framework developed in Refs.~\cite{Jujo, 2024_Kubo}, presenting an interesting direction for future research.  

A potential experimental verification of this instability could be achieved by measuring the voltage along a dc-biased superconducting wire under ac illumination.  
When the combination of ac frequency and bias current falls within the instability region shown in Fig.~\ref{fig6}, the homogeneous current-carrying state transitions into an inhomogeneous state via vortex nucleation, generating a finite voltage due to vortex motion.  
It is important to note that if the ac perturbation frequency exceeds the spectral gap (\(\hbar\omega > 2\epsilon_g\)), quasiparticle-induced dissipation may significantly increase the temperature of the superconducting wire, potentially interfering with the measurement.  
To mitigate this effect, it is preferable to focus on lower frequencies, which correspond to the region on the left side of the dashed curve in Fig.~\ref{fig6}. 
This ensures that the instability dynamics remain dominated by the Higgs mode rather than excessive quasiparticle excitations.

While preparing this paper, the author became aware of a recently published preprint~\cite{Grankin} that investigates a model in which the Higgs mode is excited through modulation of the BCS coupling.  
Their study demonstrates that the homogeneous solution can lead to a negative superfluid density, signaling an instability.  
In our work, the Higgs mode is excited by an ac perturbation superposed on a strong dc bias, whereas in their model, it is driven by BCS coupling modulation.  
Despite these different excitation mechanisms, both studies highlight the critical role of the Higgs mode in destabilizing the homogeneous superconducting state.  
Additionally, Ref.~\cite{Wang} reports that in the dirty limit, the Higgs contribution can lead to negative \( {\rm Im} \, \sigma \), in agreement with our findings.  
A more detailed and systematic theoretical investigation of this phenomenon is necessary.  
Although beyond the scope of this paper, it remains an important challenge for future research.

%%%%%%%%%%
\subsection{Kinetic inductance under a weak dc bias}
%%%%%%%%%%

Section~\ref{weak_dc_Lk} investigated the bias-dependent kinetic inductance for $J_b\ll J_{\rm dp}$, a key parameter for superconducting device applications.  
For small dc biases, \( L_k \) follows the small-bias expansion in Eq.~(\ref{Lk_expansion}), increasing monotonically with \( J_b \).  
The coefficient \( C \) for the \( {\rm ac} \parallel {\rm dc} \) (\( C \simeq 0.409 \)) and \( {\rm ac} \perp {\rm dc} \) (\( C \simeq 0.136 \)) configurations exactly matches previous results for dirty-limit superconductors, corresponding to the oscillating and frozen superfluid density assumptions, respectively~\cite{2020_Kubo, 2020_Kubo_erratum}.  
While this agreement was previously confirmed numerically in Ref.~\cite{2024_Kubo} using the Keldysh-Eilenberger theory, our results now provide an analytical verification.

The oscillating and frozen superfluid density assumptions were originally termed \textit{slow} and \textit{fast} experiment assumptions, based on whether the ac frequency is below or above the inverse relaxation time of \( n_s \).  
However, Ref.~\cite{2024_Kubo} and this work clarify that the true distinction arises not from ac frequency but from Higgs mode excitation: active in \( {\rm ac} \parallel {\rm dc} \) and absent in \( {\rm ac} \perp {\rm dc} \).

To experimentally determine the coefficient \( C \) and validate the theory, one can employ techniques such as those described in Refs.~\cite{Enpuku, Annunziata, Gao, Frasca, Makita, Dai, Greenfield}.  
It is important to note that the value of \( C \) obtained in this paper is strictly valid for the dirty limit.  
For arbitrary mean free paths, \( C \) can be computed using the Keldysh-Eilenberger theory, as shown in Fig.~8(d) of Ref.~\cite{2024_Kubo}.

%%%%%%%%%%
\subsection{Kinetic inductance under a dc bias approaching $J_{\rm dp}$}
%%%%%%%%%%

Section~\ref{strong_dc_Lk} examined the impact of stronger dc biases on \( L_k \).  
It was shown that the Higgs-induced instability leads to a divergence in \( L_k \), followed by an abrupt drop to negative values [see Fig.~\ref{fig8}(a)].  
This negative \( L_k \) signifies a transition to an inhomogeneous state, typically involving phase slips or vortex dynamics.  
In real experiments, rather than directly observing negative \( L_k \), one would measure an effective \( L_k \) corresponding to these emergent inhomogeneous states.  
Since our analysis is based on the homogeneous current-carrying solution, it does not apply to calculating \( L_k \) within the inhomogeneous regime.  
What we can conclude is that \( L_k \) diverges and then takes a finite value close to \( J_{\rm dp} \),  
which is consistent with the experimentally observed suppression of kinetic inductance near the depairing current~\cite{Enpuku, Annunziata}.  
A detailed investigation of the inhomogeneous state would provide a quantitative explanation for this phenomenon.  
Although this is beyond the scope of this paper, it remains an important challenge for future research.

Interestingly, by exploiting the \( L_k \) divergence at the boundary of the instability domain (see Fig.~\ref{fig6}), kinetic inductance can be significantly enhanced.  
For instance, Fig.~\ref{fig8}(b) shows that \( L_k \) can exceed the zero-current value by nearly two orders of magnitude well below \( J_{\rm dp} \).  
These findings indicate that, by carefully tuning the bias strength and ac frequency, the Higgs mode can be utilized to control and optimize kinetic inductance in superconducting devices.

%%%%%%%%%%
\subsection{Other implications for Superconducting Device Applications}
%%%%%%%%%%

The Higgs-induced instability renders a homogeneous high-current-carrying state susceptible to bias fluctuations or stray subgap photons, such as blackbody radiation, when \( (\omega, J_b) \) falls within the instability domain shown in Fig.~\ref{fig6}.  
This can lead to a transition into an inhomogeneous state via vortex nucleation.  
It is possible that this effect has already manifested in various superconducting applications without being explicitly recognized.

In superconducting nanowire and microstrip single-photon detectors, achieving a low dark count rate (DCR) is crucial for high-performance photon detection.  
Several known mechanisms contribute to DCR, including black-body radiation~\cite{Zadeh}, vortex crossings due to current-assisted unbinding of vortex-antivortex pairs~\cite{Yamashita}, and thermal fluctuations~\cite{Kogan}.  
While current technology has significantly reduced DCR, further studies are necessary to fully identify and mitigate the remaining sources of dark counts.  
Since these photon detectors operate under a strong dc bias close to \( J_{\rm dp} \), ac perturbations falling within the Higgs-induced instability domain shown in Fig.~\ref{fig6} may contribute to DCR.  
This potential contribution of Higgs-mode-induced instability to DCR has not been considered in previous studies and could be an overlooked factor in understanding its origins.

The ultimate accelerating gradient of superconducting cavities for particle accelerators is believed to be limited by the superheating field \( B_{\rm sh} \)~\cite{Sethna, Transtrum, Lin_Gurevich, 2017_Kubo, Wave, 2020_Kubo, 2021_Kubo, Kubo_RRR}, which represents the stability threshold of the Meissner state and corresponds to the magnetic field at which the surface current density reaches \( J_{\rm dp} \).  
Since the Higgs-induced instability weakens the stability of the current-carrying state near \( J_{\rm dp} \), it is natural to expect that the dc superheating field is also susceptible to ac perturbations.  
The onset of instability under an ac perturbation occurs at \( J_b/J_{\rm dp} \gtrsim 0.6 \), suggesting that the dirty-limit superheating field~\cite{Lin_Gurevich, 2020_Kubo, 2021_Kubo}, \( B_{\rm sh} \simeq 0.8 B_c \), may become unstable against ac perturbations at approximately \( B \sim 0.6 \times 0.8 B_c \simeq 0.4 B_c < B_{\rm sh} \).  
However, it should be noted that in clean superconductors, the Higgs mode is significantly weaker, implying that the Higgs-induced instability may not be relevant in that regime.

A superconducting diode exhibits an asymmetric critical current, allowing dissipationless current flow in one direction while suppressing it in the opposite direction.  
The system considered in this paper is particularly relevant to superconducting diodes based on simple conventional superconductors (e.g., Refs.~\cite{Vodo, Suri, 2023_Kubo, Hou}).  
If one polarity of the critical current falls within the instability domain, the superconducting state in this direction may become unstable against ac perturbations, potentially degrading the performance of the superconducting diode.  
This suggests that the Higgs-induced instability could be an overlooked factor influencing the robustness and efficiency of superconducting diodes under strong dc bias.

This instability is expected to be a ubiquitous phenomenon in superconducting applications that employ disordered superconductors under strong dc bias.  
To mitigate this instability, the most straightforward approach is to increase the mean free path of the material, which suppresses the Higgs resonance dip in \( {\rm Im} \, \sigma \) (see Refs.~\cite{Jujo, 2024_Kubo}) and effectively raises the instability threshold to higher bias currents.

Conversely, this instability could be exploited to design a new class of superconducting detectors.  
Since the instability frequency window depends on the dc bias strength (see Fig.~\ref{fig6}), 
the sensitive frequency range can be dynamically tuned via the applied bias.

%%%%%%%%%%%%%%%
%%%%%%%%%%%%%%%
%acknowledgment
%%%%%%%%%%%%%%%
%%%%%%%%%%%%%%%

\begin{acknowledgments}
I am deeply grateful to everyone who generously supported my extended paternity leave, which lasted three years~\cite{ikuji}.  
During this period, I was able to dedicate significant time to initiating various new projects and engaging in extensive email discussions with colleagues in the superconducting rf cavity and superconducting detector communities.  
These interactions and collaborations ultimately led to the development of this work.  
This work was supported by JSPS KAKENHI Grants No. JP17KK0100 and Toray Science Foundation Grants No. 19-6004.
\end{acknowledgments}

\appendix

%%%%%%%%%%%%%%%%%%%
%%%%%%%%%%%%%%%%%%%
\section{Solution of the Thermodynamic Usadel Equation at \( T \to 0 \)} \label{appendix_1}
%%%%%%%%%%%%%%%%%%%
%%%%%%%%%%%%%%%%%%%

In this appendix, we summarize the known solution for \( T \to 0 \), originally derived by Maki several decades ago~\cite{Maki, Maki_book}:
\begin{eqnarray}
&&\frac{\Delta_b(s,0)}{\Delta_0}=\exp \Bigl[ -\frac{\pi s}{4 \Delta_b(s,0)} \Bigr] , \\
&&\frac{n_s(s, 0)}{n_{s0}} = \frac{\lambda_0^2}{\lambda^2(s,0)} = \frac{\Delta_b(s,0)}{\Delta_0} -\frac{4s}{3\pi \Delta_0} , \\
&&J_b (s,0) = \sqrt{\frac{\pi s}{\Delta_0}} \Bigl\{ \frac{\Delta_b (s,0)}{\Delta_0}-\frac{4s}{3\pi\Delta_0} \Bigr\} \frac{H_{c0}}{\lambda_0}  . 
\end{eqnarray}
The depairing current density at \( T=0 \), corresponding to the maximum value of \( J_b(s,0) \), is given by
\begin{eqnarray}
&&J_{dp}(0) = \sqrt{\frac{\pi s_d}{\Delta_0}} \Bigl\{ \frac{\Delta_d}{\Delta_0}-\frac{4s_d}{3\pi\Delta_0} \Bigr\} \frac{H_{c0}}{\lambda_0}  =0.595 \frac{H_{c0}}{\lambda_0} ,\\
&& \eta_d = \frac{2}{\pi}+\frac{3\pi}{8} - \sqrt{\Bigl( \frac{2}{\pi} + \frac{3\pi}{8}\Bigr)^2 -1 } =0.300, \\
&&\Delta_d/\Delta_0 = e^{-\pi \eta_d/4} =0.790 ,\\
&&s_d/\Delta_0= \eta_d \Delta_d/\Delta_0  =0.237 .
\end{eqnarray}
In particular, for a small bias dc, the superconducting gap behaves as \( \Delta(s)/\Delta_0=1 - \pi s/4 \Delta_0 \), and the current density follows
\( J_b (s,0)= \sqrt{\pi s/\Delta_0} H_{c0}/\lambda_0 \).
This leads to the relation
\begin{eqnarray}
s = s_d \biggl( \frac{\Delta_d}{\Delta_0} - \frac{4 s_d}{3\pi\Delta_0} \biggr)^2 \biggl( \frac{J_b}{J_{\rm dp}} \biggr)^2
\end{eqnarray}
for \( T\to 0 \) and \( s/\Delta_0 \ll 1 \), which is useful for converting \( s \) into \( J_b \) in the small bias dc regime.

%%%%%%%%%%%%%%%%%%%
%%%%%%%%%%%%%%%%%%%
\section{Sanity Check: Reproducing the Results of Moor et al.} \label{appendix_2}
%%%%%%%%%%%%%%%%%%%
%%%%%%%%%%%%%%%%%%%

Our formulation and results apply to arbitrary bias dc strengths.  
However, as a consistency check, it is useful to consider the perturbative bias dc limit.  
This case was analyzed in the pioneering work of Moor et al.~\cite{Moor}, where both the ac field and dc bias were treated as perturbations.  

In this limit, we can use the zero-current Green functions $G_0 = (\epsilon + i0) / \sqrt{(\epsilon + i0)^2 - \Delta^2}$ and $F_0 = \Delta / \sqrt{(\epsilon + i0)^2 - \Delta^2}$.  
Substituting \( G_0 \) and \( F_0 \) into \( G_b \) and \( F_b \) in Eqs.~(\ref{zeta}), (\ref{kappa}), (\ref{zeta_ano}), and (\ref{kappa_ano}), we obtain
\begin{eqnarray}
&&\zeta = \frac{G_{0+}G_{0-} + F_{0+}F_{0-} +1}{\sqrt{+} + \sqrt{-}} , \\
&&\zeta^a = \frac{-G_{0+}G_{0-}^* - F_{0+}F_{0-}^* +1}{\sqrt{+} - \sqrt{-}^*} , \\
&&\kappa = -i \frac{2\epsilon \Delta (G_{0+} +G_{0-})}{\sqrt{+} \sqrt{-} (\sqrt{+}+\sqrt{-}) }, \\
&&\kappa^a= i \frac{2\epsilon \Delta (G_{0+} -G_{0-}^*)}{\sqrt{+} \sqrt{-}^* (\sqrt{+}-\sqrt{-}^*) }
\end{eqnarray}
where $\sqrt{\pm}:= \sqrt{(\epsilon+i0 \pm \hbar\omega/2)^2-\Delta^2}$. 

These expressions exactly reproduce the results obtained by Moor et al.~\cite{Moor}.  
Furthermore, the same results can be derived from the leading-order approximation of the Keldysh-Eilenberger theory in the dirty limit~\cite{2024_Kubo}.  
This provides additional validation of our approach.

\end{document}